\documentclass[prl,twocolumn,superscriptaddress]{revtex4} 

\usepackage{graphics}           
\usepackage{bm}                 
\usepackage{amsmath}            

\usepackage{verbatim}           
\usepackage{amssymb}
\usepackage{amsthm}             
\theoremstyle{plain}            
\newtheorem{theorem}{Theorem}

\begin{document}

\title{Rethinking Renormalization for Quantum Phase Transitions}

\author{Hilary A. Carteret}
\email{cartereh@iro.umontreal.ca}
\affiliation{Institute for Quantum Computing,
             University of Waterloo, Waterloo, Ontario, N2L 3G1, Canada}
\affiliation{Laboratoire d'Informatique Th\'eorique et Quantique, 
             D\'epartement d'Informatique et de Recherche Op\'erationelle,
             Pavillon Andr\'e-Aisenstadt,
             Universit\'e de Montr\'eal, Montr\'eal, Qu\'ebec, H3C 3J7, Canada}

\date{May 27, 2004}

\begin{abstract}
This is a conceptual paper that re-examines the principles underlying 
the application of renormalization methods to quantum phase 
transitions in the light of quantum information theory.  
We start by describing the intuitive argument known as the Kadanoff 
``block-spin'' construction for spins fixed on a lattice and then outline 
some subsequent ideas by Wilson and White.  We then reconstruct these concepts 
for quantum phase transitions from first principles.  
This new perspective offers some very natural explanations for some features 
of renormalization theory that had previously seemed rather mysterious, even 
contrived.  It also offers some suggestions as to how we might modify 
renormalization methods to make them more successful.  We then discuss some 
possible order parameters and a class of functionals that are analogues 
of the correlation length in such systems.  
\end{abstract}


\maketitle

\tableofcontents


\section{Introduction}\label{intro}

It is an established principle in statistical mechanics that the state of a 
system undergoing a phase transition is invariant on all length scales: its 
microscopic structure looks the same under successive changes of scale, 
and reductions in the number of degrees of freedom, known as 
``coarse-grainings''.  This is the starting point for much of modern 
statistical mechanics.  

A distinction needs to be made between the two types of phase transition 
with quantum features.  There are {\emph{quantum-to-classical}} phase 
transitions and {\emph{quantum-to-quantum}} transitions.  The latter are 
often referred to simply as ``quantum phase transitions'' and are the 
main focus of this paper.   These are a particular kind of phase 
transition that has critical points at zero temperature.  These systems are 
therefore confined to their ground states, which are usually assumed to be 
non-degenerate.  The quantum-to quantum transition behaviour usually persists 
a little way into the finite temperature regime.  When the system becomes so 
hot that the low-temperature behaviour breaks down, a quantum-to-classical 
transition will occur.

This paper is by no means the first time quantum information theorists have 
thought about the role of entanglement in quantum phase transitions, among 
them \cite{EntIndTrans,ON1,Esimple,Viv1DIsing,ZZX,QEcond,Entlength2,BCSconc,
Optlatt,FanLloyd,Guifre1,Guifre2}.  See also \cite{EntRev1,eliquids,Sachdev}.
In \cite{Dorit}, Aharonov demonstrated the existence of a quantum-to-classical 
phase transition in noisy quantum computers.  Since then a number of 
papers have used methods from statistical mechanics to understand these 
systems \cite{anyonQC,TQC,topomemory,Higgstrans} so it is well known that 
these systems fall within the larger class of phase transitions with quantum 
features.  
The aim of this paper is rather different.  We will attempt to use ideas 
from quantum information theory to develop a more general theory of 
quantum phase transitions.  

It has been suggested \cite{Nielsen,PreskillFD,Dorit} that entanglement 
should play an analogous role to the connected spin-spin 
correlation function in these systems as any correlations in a pure state 
that cannot be accounted for in terms of shared expectation values must be due 
to entanglement.  Therefore a system at a quantum phase transition should 
exhibit entanglement at all length scales.  

It is worth noting that this plausibility argument tacitly assumes that the 
ground state of the system is non-degenerate; otherwise the fact that the 
system is in its ground state would not immediately imply that the state is 
pure, and thus the fact that a reduced density matrix is mixed need not imply 
the presence of entanglement.  Nevertheless, we will assume that this 
argument is indeed correct and proceed to examine its consequences in 
section III, even though this lacuna will come back to haunt us 
later on.

Renormalization methods have been enormously successful in statistical
mechanics, but have had some difficulties when applied to quantum phase
transitions.  This paper will re-examine the conceptual foundations that 
underlie the application of renormalization methods to quantum phase 
transitions in condensed matter physics.  We will only consider systems 
of spins on lattices, where each lattice site is permanently occupied by a 
single spin.  While we hope that the ideas in this paper will be useful in 
other systems, they are beyond the remit of this paper.  

We will start by outlining the history of this approach, beginning with the 
Kadanoff picture.  
We then describe two alternative ways of thinking about renormalization.  
The starting point for both is the Kadanoff picture, which we revisit using 
ideas from quantum information theory. 

The first approach explores the consequences of the conjecture that 
entanglement plays an analogous role in quantum phase transitions to that 
of correlation functions in classical phase transitions.  This implies that 
the state of a system undergoing a quantum phase transition should exhibit 
entanglement at all length scales.  This idea is intuitively fruitful, but it 
does not seem to be possible to use this to obtain quantitative results 
using existing techniques.  

The second approach starts by considering the interactions in 
the system, and leads to a new way of thinking about numerical renormalization 
methods which can explain both their previous successes and failures.  This 
suggests both novel strategies for modifying these numerical renormalization 
techniques and new physical insights into the fundamental principles 
underlying the application of renormalization to condensed-matter systems.  
We finish by discussing what this implies about the order parameters of the 
system and define a class of functionals that are possible candidates 
to replace the correlation function in such systems.

The only initial assumptions we will make about the Hamiltonian is that 
it is dominated by short-range interactions which involve a small number 
of spins (which may be more than two).  It can be shown that pairwise 
interactions can generate four-spin interactions in physically reasonable 
systems \cite{4dot,3-4HamQC}.  We must also assume that the 
Hamiltonian's structure repeats on a scale that remains a constant as the 
infinite lattice limit is taken, as short-rangedness on its own is not 
enough to guarantee that the problem will be tractable \cite{QMAcompLH}.

\section{I: Renormalization revisited}\label{revisit}

In order to explain what we're doing and why, we will go right back to the 
origins of renormalization theory in statistical mechanics.  This had its 
origin in some observations in classical statistical mechanics.  Consider 
a system that undergoes a phase transition at a certain temperature, such as 
a chain of spins in a ferro-magnetic material.  
If we coarse-grain a system that is below the critical temperature, using 
some kind of averaging procedure (it generally doesn't matter which) then 
the system will generally look a lot like itself afterwards, only at a lower 
temperature.  Likewise, if we try this averaging procedure with a system 
above the critical temperature, we find that the coarse-grained looks hotter.  
What happens in the middle?  This is where the phase transition occurs.  For 
this special temperature, the coarse-grained system appears to have exactly 
the same temperature after this averaging procedure; it is invariant under 
rescaling. It is this entirely phenomenological observation that originally 
inspired the application of renormalization theory to statistical mechanics.

\subsection{I.1: The Kadanoff construction}\label{Kadanoff}

The first attempt to formulate a microphysical theory to account for this 
observation was by L.P. Kadanoff \cite{Kadanoff1,Goldenfeld}.  His insight 
began with the realisation that the fact that a system at criticality looks 
the same at all length scales means that its dynamics must also be 
self-similar.
How can a system made of spins on a lattice do this?  Kadanoff's proposal 
was that small collections or ``blocks'' of spins must behave collectively 
like single spins.  He called these emergent spins ``{\bf{block-spins}}'' and 
suggested that these will also team up to behave like a single emergent spin, 
and so on.   This hierarchical structure should continue over distances less 
than the correlation length of the system.  
Likewise, the effective Hamiltonian coupling those emergent spins should 
have the same functional form as that between the original physical spins.   
So he divided up the lattice into little blocks of $L$ spins and postulated a 
recursive procedure (called a ``{\bf{block-spin transformation}}'') to map 
the spins in each block into their single, collective spin.   

\begin{figure}[floatfix]
    \begin{minipage}{\columnwidth}
    \begin{center}
        \resizebox{0.8\columnwidth}{!}{\includegraphics{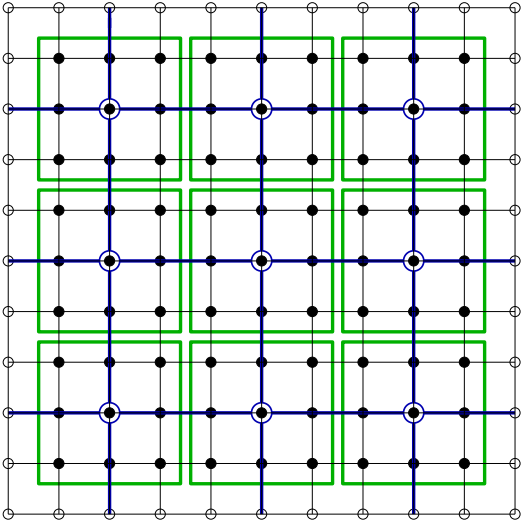}}
    \end{center}
    \end{minipage}
    \caption{A Kadanoff ``blocking'' for a square lattice.  The original 
     lattice (black) is divided up into blocks (green/grey squares) and a 
     single, collective ``block spin'' is defined for each block (blue/black 
     rings). A rescaled lattice of block spins is obtained (thick blue/black 
     lines) which has the same geometry as the original lattice.}
    \label{heuristic}
\end{figure}

These block-spins would in turn give rise to emergent spins, coupled by their 
own effective Hamiltonian, and so on {\emph{ad infinitum}}.  
Figure~\ref{heuristic} contains a heuristic picture of this construction for a 
square lattice.  
Unfortunately, while his intuitive argument seemed to account for the Widom 
scaling laws \cite{Widom}, it had a number of features that made little 
physical sense.  

\subsection{I.2: Wilson renormalization}\label{Wilson}

Wilson reworked this idea in two papers in 1971 \cite{Wilson1,Wilson2}.  
He began the first of these two papers by describing the ``Kadanoff picture'' 
and pointing out why it didn't make any sense, 
\newline

{\emph{``The idea that blocks of spins act as a unit near the critical 
temperature does not stand up to close examination; in fact only near zero 
temperature [\ldots] is it true \ldots}}

{\emph{\ldots In short the Kadanoff block picture, although absurd, will be 
the basis for generalizations which are not absurd, and it is helpful to 
understand the Kadanoff picture in differential form \ldots''}}
\newline

He makes this point even more strongly in the second paper:
\newline

{\emph{``No justification has been found for the Kadanoff picture.  The 
difficulty is that Kadanoff assumes that the effective spin variable $s_m'$
has only two values, up or down.  In the exact Ising model a block of spins
of size $L$ and $z^{L^d}$ configurations (where $d$ is the dimensionality of 
the system) and it is hard to see how to reduce the $z^{L^d}$ different 
configurations to just two.  Because there has been no justification of the 
Kadanoff picture, it has been impossible to calculate specific critical 
exponents within the Kadanoff picture \dots}}
\newline

He then extensively modified Kadanoff's block-spin rescaling idea.  Having 
just explained why it doesn't work for integer values of $L,$ he argued that 
Kadanoff's picture should be interpreted as an analyticity assumption, 
and that this {\emph{was}} correct.  
He then proceded to assume that Kadanoff's picture not only worked when 
averaging over blocks of integer size (to obtain a new block of size $nL$ 
in the original lattice spacing) but that it also worked for 
$L \mapsto (1+\delta)L.$  
This enabled Wilson to write the Kadanoff-Widom scaling laws in differential 
form and obtain a well-defined renormalization procedure.  
This led to a method to calculate the critical exponents for the transition 
using the behaviour of the system when the renormalization procedure is 
iterated.   In his paper ``II'' \cite{Wilson2} he successfully applied this 
methodology to the Kondo problem.  Note also that Wilson's renormalization  
implicitly assumes a Kadanoff block-like structure in the vicinity of the 
critical point, as well as at the critical point itself. 
 
The so-called ``standard real-space RG approach'' \cite{WNoack92} consists 
of dividing a lattice into a set of blocks (assumed to be the same) and 
diagonalizing each block locally to find a set of eigenstates.  This set is 
then truncated to leave $m$ states which are deemed to be the ``most 
important'' (where $m$ is a parameter that we are free to choose) and the 
reduced set is used as an eigenbasis to construct an approximate Hamiltonian 
$H_B$ for a new, larger block, made by merging {\emph{two}} of the original 
blocks.  All other operators are likewise ``renormalized'' by projection 
onto this $m$-dimensional subspace.

The original version of this idea completely neglected interactions between 
the blocks.   When subsequent attempts to  use this technique on other systems 
got into difficulties, Wilson suggested using degenerate perturbation theory 
to try and take account of the interactions between blocks, but this 
modification didn't seem to work very well \cite{WNoack92}.

\subsection{I.3: The density matrix renormalization group}\label{DMRG}

Wilson's numerical approach worked well for impurity problems, but not 
much else \cite{White92}.  In 1992 - 93, White (and Noack) defined a family 
of alternative renormalization procedures \cite{WNoack92,White92,White93algs}, 
of which the most successful have come to be known collectively as density 
matrix renormalization group methods (DMRG).  All of these various approaches 
were attempts to take better account of the interaction of the block with 
its surroundings.  This is the reason for choosing the basis for the 
rescaled system using the state instead of the Hamiltonian, as explained by 
White in \cite{White93algs}:
\newline

``{\emph{...in reality the block is not isolated, the density matrix is not 
$exp(-\beta H_B)$ [...] and eigenstates of the block Hamiltonian are not 
eigenstates of the block's density matrix.  For a system which is strongly 
coupled to the outside universe, it is much more appropriate to use the 
eigenstates of the density matrix to describe the system rather than the 
eigenstates of the systems's Hamiltonian.}}'' 
\newline

A rederivation of White's procedure in terms of the Schmidt decomposition of 
density matrices and entanglement is given in \cite{ON1}.  The essential idea 
of the DMRG is to take {\emph{two}} blocks and form a ``superblock'', which 
is then diagonalized.  If the superblock is in a pure state, this will 
simultaneously diagonalize the two subsystem (block) reduced density matrices. 
Then a basic result from linear algebra is used (by both White, and Osborne 
and Nielsen) to show that the optimal choice of states is to choose the $m$ 
largest eigenvalues of the block density matrix. 

Note that both these approaches think of the block-spin transformation used to 
rescale the system as an averaging process; Wilson treats it in terms of some 
kind of ``mean'' spin, whereas the DMRG is more like taking the ``mode''.  
The leitmotif that unites all these approaches seems to be that the essential 
step in defining a workable renormalization procedure is to specify a 
block-spin transformation and be able to justify it.

\subsection{I.4: The renormalization group flow}\label{critsurface}

\begin{figure}[floatfix]
    \begin{minipage}{\columnwidth}
    \begin{center}
        \resizebox{0.8\columnwidth}{!}{\includegraphics{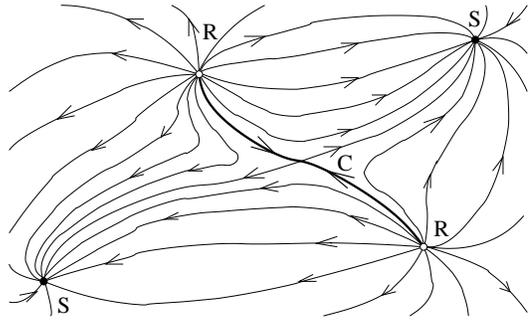}}
    \end{center}
    \end{minipage}
    \caption{A heuristic diagram of a renormalization group flow, which 
             follows the arrows as shown.  The points marked ``S'' are 
             attractors, or ``sink'' points and can be thought of as local 
             minima.  Those marked ``R'' are repulsive, or completely 
             unstable points, rather like maxima.  The point marked ``C'' is 
             a critical point.  These are quasi-stable when approached in 
             some special directions (thick line) and unstable from others, 
             rather like saddle points, and are the focus of this paper. 
             The thick line is the critical surface, or manifold.}
    \label{RNGflow}
\end{figure}

All of these methods require that the renormalization procedure is iterated. 
If we record the rescaled states and effective Hamiltonians at each step 
and plot these in their respective parameter spaces, we see some very 
distinctive patterns, such as those in figure~\ref{RNGflow}.  The motion 
of the state (or Hamiltonian) under iteration of the renormalization 
procedure is known as the renormalization group flow, or RNG flow. 
 
A characteristic feature of the critical points under state renormalization 
is that they are associated with a set of states (often called the critical 
``manifold'' or ``surface'') which consists of a lower-dimensional set of 
neighbouring states which are drawn into the critical point under the 
renormalization group flow, in the limit as the number of iterations of the 
renormalization group tends to infinity.  This manifold is represented by 
the thick line in figure~\ref{RNGflow}.  The fixed points of this iteration
are the only places in the parameter space in which the state is exactly 
renormalization invariant.   System parameters (state or Hamiltonian) that 
drive it away from critical points are known as ``relevant'' parameters;  
those whose initial values the system tends to forget over time are known as 
``irrelevant'' parameters. 

Wilson made this point in his first paper \cite{Wilson1} in which he 
compared the parameter space to a landscape, with hills as the replusive 
points and valleys leading to sink points.  Critical points corresponded 
to saddles, or ``passes'' between hills.  This is how the analytic behaviour 
of a finite system can nevertheless produce discontinuous behaviour in the 
infinite lattice limit; think of a ball rolling down either side of 
a hill.  At finite times it may be part way down, but after an infinite 
amount of time it must be sitting at the bottom of some kind of minimum.  
(It is just such an infinite-lattice limit discontinuity that resolves the 
apparent paradox in \cite{Goldenfeld}, which uses time reversal symmetry to 
``prove'' the impossibility of phase transition, this being precisely the 
point Goldenfeld is making.)

\subsection{I.5: Entanglement and quantum phase transitions}\label{philosophy}

In \cite{ON1} the authors argued that the failures of existing 
renormalization schemes to find the fixed points corresponding to quantum 
phase transitions might be caused by the the fact that they 
either completely fail to take account of the role of entanglement, or do 
so only imperfectly.  The authors then went on to propose a modification 
of the density matrix renormalization group (DMRG) to optimize the 
preservation of the entanglement of formation between successive steps of 
the renormalization procedure.  They were unable to give a closed form for 
this procedure, because there is no known general closed form for the 
entanglement of formation.  (Indeed, as the bipartite separability 
{\emph{decision}} problem has since been proved to be NP-{\small{HARD}}
\cite{Gurvitsep}, such a general closed form is unlikely to ever be found.) 

This lack of closed forms has caused most previous papers that have examined 
the entanglement properties of specific systems to take at least one of two 
approaches.  The first of these is to use the Wootters formula for the 
concurrence for two qubits \cite{concurrence} to calculate the entanglement 
of formation between pairs of spins, see for example \cite{Viv1DIsing}.  
The other approach is to look at the entanglement between blocks of spins 
and the rest of the lattice, such as in \cite{Guifre1,Guifre2,finegloss}.
The results in \cite{Guifre1,Guifre2} led the authors to suggest an 
alternative explanation for the difficulties encountered with the DMRG: they 
suggested that the cause of the convergence problems was that at criticality, 
there weren't enough insignificant subspaces which could be safely discarded 
without disrupting the qualitative behaviour of the state.

\section{II: The Quantum Kadanoff Construction}\label{Quantanoff}

We will now go right back to the first principles of renormalization and 
rework Kadanoff's construction.   
We will begin by dividing the lattice up into blocks of $n$ spins each, 
and assume that the spins in the block are in some state that enables 
them to behave as a single, quantum spin.  
(This state might be entangled or it might not: this construction does not 
depend on entanglement.)
For simplicity, we will assume that each block generates just a single spin; 
an assumption that we will justify later.  
We would like to find some procedure that would enable us to implement the 
transformation from the physical spins to the emergent block-spin.  

Let $\mathcal{B}$ be the transformation from the constituent spins to the 
block-spin, defined on a single block.  The collective transformation for 
the entire lattice, $\mathcal{R},$ is obtained by performing $\mathcal{B}$ 
on all the blocks for some tiling over the entire lattice.   I will use 
the superscript $(r)$ to denote the spins and block-spins of the system after 
this procedure has been iterated $r$ times.  

The renormalization transformation is a way to change between the frame of 
reference of the physical spins, and that of the block spins, and its 
functional form should reflect that underlying dynamics.  
It may therefore be unwise to define a renormalization procedure 
{\emph{a priori}} and then look for its fixed points.  
Thus we will {\emph{not}} define an explicit $\mathcal{B},\mathcal{R}$ at 
this stage, but instead attempt to find a natural way to determine what 
these transformations should be.  

Let $\rho^{(r)}$ be the state of a single $r$th-order block, obtained by 
tracing out the rest of the lattice, and let 
\begin{equation}
 \chi_i^{(r)} = {\rm{tr}}_{1,\ldots,i-1,i+1,\ldots,n}\rho^{(r)}
\end{equation}
be the state of the $i$th single spin from that block.  As $\mathcal{R}$ is 
defined blockwise it will commute with the trace for subsystems consisting of 
intact blocks, so 
\begin{equation}\label{recursion}
 \chi_i^{(r+1)} =  \mathcal{B}\rho^{(r)} 
\end{equation}
for some $i \in 1,\ldots,n.$  
The assumption that all the blocks look the same under the iteration of the 
procedure gives us that
\begin{align}\label{renorm2}
 &\rho^{(r+1)}=\rho^{(r)}, \quad &\chi^{(r+1)}=\chi^{(r)}
\end{align}
Where we allow that the local density matrix $\rho^{(r)}$ may be mixed for 
any finite $r.$  
The assumption that the state of the block spins after rescaling looks 
the same as the individual spins before rescaling can be written as
\begin{equation}\label{intrablock}
  \mathcal{B}\rho^{(r)} = \chi_i^{(r)} = 
  {\rm{tr}}_{1,\ldots,i-1,i+1,\ldots,n}\rho^{(r)},
\end{equation}
where we have implicitly assumed a weak form of homogeneity: that the spins 
all have the same local density matrix.

\bigskip

It is common practice in statistical mechanics to include a projection 
operation, followed by multiplication by a scalar greater than $1$ to 
``renormalize'' the state as part of such coarse-graining procedures.  This 
is rather disturbing to a quantum information theorist \cite{Guifre,monotone}, 
not least because there is a real risk that such ``free'' post-selection will 
alter the energy-density of the system, just as ``collapse model'' 
interpretations of quantum mechanics tend to have serious problems with 
energy conservation.  I will take the position that decimation procedures 
should ultimately be interpreted passively: we are just choosing to ignore 
some degrees of freedom as being unimportant.  I will therefore disallow 
free post-selection for the rest of this paper, as I consider it to be 
unnecessary and undesirable.   
I will assume that this map consists of some change of basis, followed by 
the tracing out of those degrees of freedom that we wish to ignore. 
It will sometimes prove to be convenient to use post-selection as a 
conceptual tool in some intermediate steps in this paper but it will 
ultimately be eliminated.  

\bigskip

There are two defining features of the Kadanoff construction.  One is the 
claim that each cluster of spins behaves like a single spin.  In order to 
derive a fully quantum version of this idea we must incorporate a fundamental 
difference between a classical and a quantum spin.  For a block to give rise 
to a classical spin, it need only support the eigenstates of that spin.  
A truly {\emph{quantum}} spin requires more than this: the block must also 
support all linear combinations of those basis states as well.  
If the block fails to do this, the spin it generates will be quasi-classical.

The second feature is the assertion that the block-spin transformation is a 
recursive symmetry of the system.  This assumption goes right back to the 
original phenomenological observation of scale invariance at a phase 
transition.  Each spin, whether physical or emergent, will interact with its 
surroundings which it will perceive as its environment. 

The two statements above together beg the question of whether the block spin 
is more or less strongly coupled to its environment than the constituent 
spins from which it is made? 
Clearly, the answer to this question must be, ``sometimes more, sometimes 
less''.  If the answer is ``more'', then each block-spin will be subject 
to progressively more decoherence, until it ceases to be anything like 
a quantum spin, and the whole block structure is overwhelmed by disorder.  
This must be a sink point. 

\bigskip

If the answer is ``less'', something more interesting happens.  Recall that 
we are discussing a {\emph{fixed}} point of an iteration.  So, if the 
first layer of block-spins is less strongly coupled to its environment, then 
the next layer must also be less strongly coupled in turn, by the same 
factor, or else we are not at a fixed point of the RNG flow.  The only way 
adding a further layer of decoupling can leave the new block-spin's 
interaction with its environment {\emph{unchanged}} is if it was already 
{\emph{completely}} decoupled from that environment.

\bigskip

Clearly, there must also be a boundary between these two basins of attraction, 
where the strength of coupling is exactly the same after each iteration.  
States exactly on this boundary can neither reach the perfectly decoupled 
point nor escape to the sink and might seem to be trapped in the boundary 
surface, but this midpoint must be unstable; any fluctuations will generally 
tip it into the basin of attraction for the sink point.  Such fluctuations 
can either be thought of as a change in the state, or else as an additional 
term in the Hamiltonian, the effect will be the same.  The fluctuations will 
act as an additional kick to the system, which it cannot tolerate.
Thus the probability of finding the system in such a boundary state in the 
infinite lattice limit is zero.  
(Strictly speaking, it's not impossible that such fluctuations could tip the 
system into the ``perfectly decoupled'' basin, it's just very unlikely.)
The continuum of possibilities available to finite sized systems has 
converged to just two in the infinite lattice limit.  
It can also be seen that if such fluctuations are sufficiently strong, 
they can overwhelm even the ``perfectly decoupled'' fixed point.  Thus 
this fixed point can be only partially stable: it is a critical point.

\section{III: Entanglement on all length scales?}\label{allscales}

In this section we will assume that the conjecture that systems undergoing 
quantum phase transitions exhibit entanglement ``on all length scales'' is 
correct and see where it leads us.  For now, our goal is to build some 
intuition for the problem; we will make some of the ideas inspired by this 
more precise in a later section of this paper.

When trying to compare the role of entanglement in quantum phase transitions 
with that of classical correlation functions in more conventional phase 
transitions, it is important to note a few differences between the two. 
The quantitative study of entanglement arguably began with Bell's theorem 
\cite{Bellthm}.   This paper demonstrated a measureable difference between 
the predictions made by quantum mechanics and those made by a local hidden 
variables model.  What is of interest for our purposes here is that these 
differences are only detectable when one compares the results from 
measurements made in more than one basis.  If we restrict the allowed 
measurements to those in only one basis, a local hidden variables model 
can be constructed to account for what we see. 

It has been noted elsewhere both that entanglement cannot be shared 
arbitrarily, unlike classical correlations \cite{3tangle} and that it is 
typically competitive.  It should also be noted that there are many types 
of entanglement and in the case of multipartite entanglement, the number 
of different types grows exponentially in the number of parties 
\cite{mschmidt}.  Furthermore, different types of entanglement cannot always 
be interconverted \cite{interconv}. 

The only point we are trying to make at this stage is that we should not 
assume that the behaviour of the entanglement in a system will necessarily 
mimic the behaviour of the classical correlations in analogous-looking 
systems, or even the {\emph{same}} system.  In fact, a system has been found 
that exhibits a divergent ``entanglement length'' whilst having a finite 
correlation length \cite{Entlength2}. 
This should be a particular concern if we are using a renormalization 
procedure that includes a change of basis, such as the DMRG. 

We will argue that it is possible to bypass some of the problems caused by 
the lack of tractable entanglement measures by thinking about these systems 
in terms of their information dynamics.  We will begin by assuming the 
existence of {\emph{some}} kind of renormalization procedure and that it will 
take the form of an iterated rescaling.  

If a state is entangled with some larger system, it cannot be in a pure local 
state.  (Whether or not the converse of this statement is also true depends 
on your preferred interpretation of quantum mechanics.  The reader may regard 
thermal mixtures as being entangled with external field degrees of freedom 
if they wish; it will make no difference to the arguments in this paper.) 

We will look at the entanglement within (mixed) subsystems, rather than 
between the subsystem and the whole.  This approach has some similarites 
to the conclusion reached at the end of \cite{ON1}, but with an important 
difference: we are thinking in terms of the entanglement {\emph{within}} a 
block which we are assuming to be in a mixed state.  
By contrast, the modified DMRG suggested in \cite{ON1} is formulated in terms 
of the entanglement between the block and the ``superblock''.  
There are various reasons for this modification.  One of the problems with 
only looking at the entanglement {\emph{between}} a block and the rest of 
the system is that it implicitly requires you to assume that the system is 
globally pure, as there is no way that one can distinguish between ``proper'' 
and ``improper'' mixtures by looking at the local state alone.  
This is likely to cause problems when we try to extend such analyses to 
systems at finite temperature.  It also makes it a little difficult to talk 
about the scales over which entanglement is present in a well-defined way; 
entanglement between the block and the rest of the lattice is implicitly 
global. 

If instead we think in terms of the entanglement {\emph{within}} a block 
at each iteration, we are not neglecting the entanglement between blocks, 
as at the next step in the iteration at least some of that will become 
``internal'' entanglement.  We shall see later (in the interaction 
formulation of renormalization) that the entanglement between the block and 
its environment will have a vital role to play.  But for the purposes of 
defining the entanglement properties of critical points, we claim it makes 
more sense to think in terms of the entanglement within the block at each 
stage of the iteration. 

We will attempt to make sense of the phrase ``the state has entanglement on 
all length scales'', in this context.  In particular, we will allow for the 
possibility that this entanglement may be irreducibly multipartite.  
The lack of closed forms for mixed state multipartite entanglement will mean 
that we cannot quantify that entanglement, but we may still be able to answer 
at least some questions of interest. 
This will, of necessity, result in a difference in emphasis in the role 
attributed to entanglement (compared with that in \cite{ON1,Guifre1,Guifre2}).

\subsection{III.1: Entanglement and renormalization}\label{methodology}

We will begin by clarifying what we mean when we say ``entanglement on all 
length scales.''  
The Kadanoff construction asserts that both the state and the Hamiltonian 
have a self-similar structure, and we have concluded that the critical points 
correspond to the case when the block-spins are more decoupled from their 
environment than their constitutent spins.  We will look for entanglement 
in the context of this situation, i.e., states which have entanglement 
properties that cause them to behave in this way.  
The inequivalence of different kinds of entanglement mean that the choice of 
block structure may matter for lattices of {\emph{quantum}} spins.  In other 
words, the way in which those blocks generate their respective block spins 
is not just an averaging procedure that can be chosen arbitrarily, it 
represents an essential feature of the dynamics of these systems that gives 
rise to that emergent spin. 
We will try to find states with the right entanglement properties and 
let these lead us to the renormalization procedure and the Hamiltonian. 

Suppose that these emergent spins are themselves entangled in a way that has 
the same form of entanglement as that for the physical qubits, and that the 
rescaling procedure can be repeated on these emergent spins to obtain another 
entangled system and so on, {\emph{ad infinitum.}}  Then the system must have 
finite subsystems on all scales that are entangled, where the notion of 
``scale'' refers to the associated renormalization procedure. 
Thus the block-spin must be a non-local structure and the entanglement of 
these states will have a branching, modular structure.  

\bigskip

This feature is remarkably like one way of characterizing quantum 
error-correcting codes.  
In \cite{ThQECCs}, the authors give a number of ways of defining codes.   
Suppose the ``error'' (for which read, the difference in the state 
$\rho'$ from the code state, $\rho_c$) is of the form 
\begin{equation}
  \rho'=\mathbb{S}(\rho_c) = \sum_a A_a \rho_c A_a^{\dagger},
\end{equation}
and let $\mathcal{A}=\{A_i\}$ be the set of Kraus operators for the error 
superoperator $\mathbb{S}.$ 

\begin{theorem}{\bf{(Knill and Laflamme) 
\newline
[Theorem 3.5 in \cite{ThQECCs}]}}\label{KLisodef}
\newline
The subspace $\mathcal{C}$ of $\mathcal{H}$ is an $\mathcal{A}$-correcting 
code iff there is an isomorphism $\sigma: \mathcal{H} \simeq
\mathcal{C}\otimes\mathcal{E}\oplus\mathcal{D}$ such that for all 
$A_a \in \mathcal{A}$ and $|\Psi\rangle \in \mathcal{C}, \; 
A_a|\Psi\rangle = \sigma (|\Psi\rangle\otimes|\mathcal{E}(a)\rangle)$ for 
some vector $|\mathcal{E}(a)\rangle$ depending on $A_a$ alone. 
(For a {\emph{perfect}} quantum code, $\mathcal{D}$ is empty, and the 
vectors $|\mathcal{E}(a)\rangle$ span $\mathcal{E}.$)
\end{theorem}

Note that although the original formulation of the theory of quantum error 
correction codes was formulated {\bf{actively}} in terms of projective 
measurements followed by correction and decoding, the authors realised 
that what really matters is the isomorphism in theorem~\ref{KLisodef}, and 
that the whole structure could be understood in terms of {\bf{passive}} 
transformations.

We can now write down some states with the requisite entanglement properties, 
provided we can solve a packing problem that depends only on the geometry 
of the lattice.  In \cite{concat}, the authors showed how to construct 
entangled states whose entanglement has precisely this branching property.  
These states are concatenated quantum error correction codes.
A quantum error correction code is a highly entangled state that protects a 
quantum state from local noise.  In a single encoding, the logical 
qubit that we wish to protect is encoded into several qubits, to form a 
codeblock.  This encoding can then be concatenated: each constituent spin 
is itself encoded in its own personal code block.  This structure can then 
be repeated as many times as we like. 

Now consider this construction in reverse, from the lowest level of encoding 
to the highest.  Suppose that the lowest level of the encoding uses the 
physical qubits of the lattice.  If we arrange the qubits of a codeblock into 
a tile, we can tile the lattice with them, and the encoded logical qubits 
of the error-correction code will be the block spins.  
When we repeat the block-spin construction (concatenate the encoding) the 
logical qubits of the next level of encoding will be the new block spins, and 
so on.   If we specify that each level of encoding is performed using the 
same code, then we will obtain a state with precisely the kind of modular 
entanglement structure we were looking for.  
The corresponding renormalization procedure will just be the decoding 
operation for that code, which will leave the block in the state
\begin{equation}\label{ancilla}
 \mathcal{B}\rho^{(r)} = \chi^{(r+1)}\otimes 
            |\underbrace{00\ldots0}_{n-1 \text{ spins}}\rangle 
            \langle 00 \ldots 0 |
\end{equation}
and the degrees of freedom that we discard will be the $n-1$ ``ancilla'' 
spins.  

\bigskip

{\bf{Remark:}} It could be objected that equation \eqref{ancilla} doesn't 
tell us very much, as this statement will be true for almost any choice of 
unitary transformation if the state is locally pure.  
This is a quantum version of one of Wilson's objections to the Kadanoff 
picture, and it would be particularly apt if we were to assume that the 
state of the block is locally pure.  We are not making this assumption, 
but this is still an issue and we will return to it later in this paper.  
For now, we will call this the ``{\emph{arbitrary basis change problem}}'' 
and leave it open, though as we shall see in section IV of this paper, it 
is solved ``for free'' in the interaction formulation in section IV.

\subsection{III.2: A toy example}\label{toy}

\begin{figure}[floatfix]
    \begin{minipage}{\columnwidth}
    \begin{center}
        \resizebox{0.8\columnwidth}{!}{\includegraphics{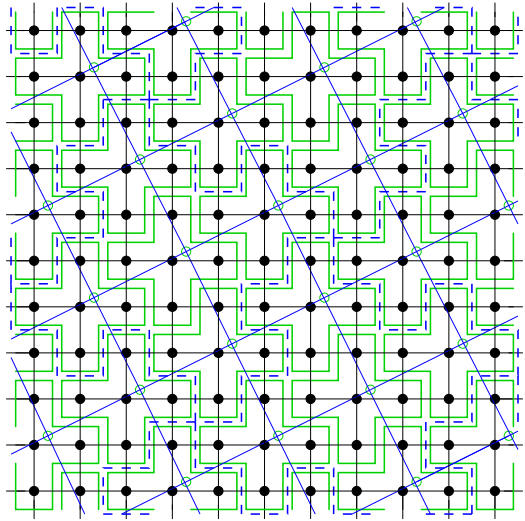}}
    \end{center}
    \end{minipage}
    \caption{A renormalization procedure for a 2-D square lattice, 
    employing the $5$-qubit perfect code and a right-handed tiling.  
    The original lattice is drawn in fine lines, the $5$-qubit code 
    blocks are the plus shapes drawn in green/grey lines, the meta-block 
    in dashed lines.  The block-spins are represented by open circles and 
    the rescaled lattice is drawn in diagonal (blue) lines.  This 
    renormalization rescales the lattice by a factor of $\sqrt{5}$ and 
    rotates it by $+\arctan(1/2).$}
    \label{2D5qR}
\end{figure}

To see how this works, here is a toy example.  It is a rather unphysical one, 
as the state is not translation invariant (amongst other things) so it 
should not be taken too seriously.  However, it does have a particularly 
transparent renormalization procedure and so should serve to help develop 
our intuition.  Figure~\ref{2D5qR} is an exact 
covering of the two-dimensional square lattice using the $5$-qubit perfect 
code from \cite{perfect5q,5qubit4}.  This is an example of an additive code.  
These codes are characterized in terms of their stabilizers; the actual choice 
of code-space is fundamentally arbitrary.  Let us fix our choice of logical 
basis states for this code to be the one in \cite{Gottesman} and then we 
can write:
\begin{multline}\label{5qubitc0}
 |0_L\rangle = \frac{1}{4}[
              |00000\rangle + |10010\rangle + |01001\rangle + |10100\rangle \\ 
            + |01010\rangle - |11011\rangle - |00110\rangle - |11000\rangle \\ 
            - |11101\rangle - |00011\rangle - |11110\rangle - |01111\rangle \\
            - |10001\rangle - |01100\rangle - |10111\rangle + |00101\rangle]
\end{multline}
The logical $|1_L\rangle$ can be obtained from $|0_L\rangle$ by interchanging 
the $1$'s and $0$'s.  This stabilizer code uses five qubits to protect 
one logical qubit.
The corresponding renormalization procedure is defined in terms of blocks of 
five spins, which behave like one spin \cite{ThQECCs}.  

{\subsubsection{Some remarks about tilings...}}

Not every code will work for every lattice system, because the packing 
problem is non-trivial for lattices of dimension two or higher.  
We can embed the code in tile(s) of any shape, so long as we use the 
same shape(s) for the whole lattice.  We must also obtain a new lattice 
that differs from the old one by only a scale factor (and possibly a 
rotation) when we perform the renormalization.  
The $5$-qubit code tiling in this example is also handed: there are two 
different ways of putting the tiles together, one of which generates a 
renormalized lattice that is rotated by $+\arctan(1/2)$ and the other by 
$-\arctan(1/2).$ 

Whether or not a given tiling generates a rescaled lattice is a property 
of the tiling alone.  
Note that for our $5$-qubit example, the choice of tiling is degenerate.
This is one of the reasons why this example is probably unphysical: it is 
not translation invariant.
In fact there are some codes which are translation invariant (under integer 
multiples of the lattice spacing) but their block-spin structure is not 
obvious, which is why I did not use them as a running example.  We will 
look at these in a later section, and show that they can be described in 
this way.  
There is also no particular reason to believe that those translationally 
invariant codes are the most general class of states with these properties, 
because they are also exact codes. 
Exact codes have much stronger convergence properties than are required, 
and are therefore somewhat over-engineered as critical points.  

A more realistic choice of code might be some kind of an approximate 
quantum code \cite{ApproxQECC,Dan-AppQECC}, which are only required to 
protect the original state with high fidelity at each encoding, instead of 
exactly.  
Unfortunately, we do not yet know necessary and sufficient conditions for 
an encoding to be an approximate error-correction code; we only know what 
these conditions are for exact codes.  This gap in our knowledge is not just 
a problem when we try to write down states: it also makes it difficult to 
discuss the renomalization group flow for these systems.  Therefore we will 
return to our rather unphysical, exact concatenated code example for the rest 
of this section and return to this question in section IV.

\begin{figure}[floatfix]
    \begin{minipage}{\columnwidth}
    \begin{center}
        \resizebox{0.8\columnwidth}{!}{\includegraphics{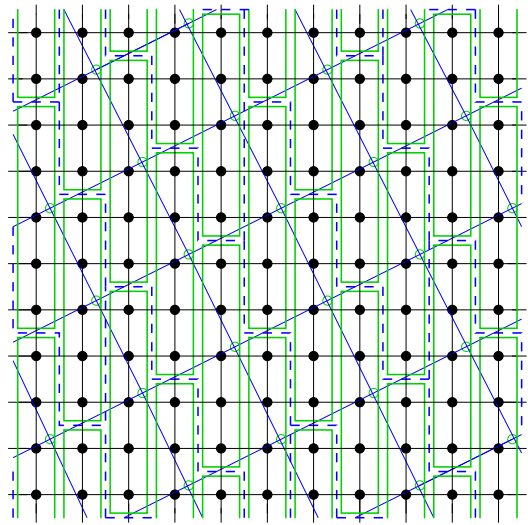}}
    \end{center}
    \end{minipage}
    \caption{An alternative tiling for the 2-D square lattice, using the 
    $5$-qubit perfect code and a right-handed tiling.  
    The original lattice is drawn in fine lines, the $5$-qubit code 
    blocks are the plus shapes drawn in green/grey lines, the meta-block 
    in dashed lines.  The block-spins are represented by open circles and 
    the rescaled lattice is drawn in diagonal (blue) lines.  This 
    renormalization also rescales the lattice by a factor of $\sqrt{5}$ and 
    rotates it by $+\arctan(1/2).$}
    \label{brick}
\end{figure}

This code-tiling construction is highly non-unique.  The same code can also 
be used in to tile this lattice in a $5 \times 1$ brick tiling, as shown 
in figure~\ref{brick}.  Note that this can be chosen to produce exactly the 
same rescaled lattice as the tiling in figure~\ref{2D5qR}.  
It is also possible to tile the lattice in figure~\ref{2D5qR} using other 
codes, such as Shor's $9$-qubit code \cite{ShorCode}.  The same is true for 
other lattices as well: a triangular lattice can be tiled using the Steane 
$7$-qubit code \cite{Steane7}, or other $7$-qubit codes \cite{MBR7}. 

\subsubsection{The renormalization flow for the state}\label{stateflow}

Theorem~\ref{KLisodef} implies that any state that differs from the 
concatenated code state by the equivalent of a correctable error moves to 
the fixed point under the corresponding renormalization: the effects of the 
local physics on these states is confined to a subspace.  
The detect-and-correct parts of the protocol merely return the system to some 
standard form for our convenience when reading out the state.  This is 
essentially a relabelling, and the physics of the system should not depend 
on our choice of basis labels for the block-spins.  As such, we can arguably 
interpret the measurement and conditional operations {\emph{passively,}} as 
something that we bring to the system as human beings trying to analyse it, 
rather than an {\emph{active}} interpretation in terms of something that we 
would need to actually do to the system.  

In a concatenated code, if a small number of lower-level codeblocks fail, 
then the larger encoding can still cope but there is a limit to what errors 
it can correct, even if we are allowed to use arbitrarily many levels of 
encoding.  This ``failure'' can be interpreted as the boundary of the basin 
of attraction for this critical point under our renormalization procedure.  
States just inside the boundary will eventually reach the code state 
(critical point) under the error-correction procedure.  States just outside 
that boundary will move off towards some other fixed point because the 
error-correction procedure is unable to move them towards the code state.  

It can now be seen that the difference between a ``correctable'' and a 
``fatal'' error in the {\emph{concatenated}} error correcting code will be 
closely related to the difference between the ``irrelevant'' and ``relevant'' 
parameters under the renormalization group flow.  (Errors that the code can 
only {\emph{detect}} must count as failures.)  Eventually, any state that 
differs from the fixed point state by such a relevant parameter will be driven 
away under the action of the renormalization group.   
States exactly on the boundary of the basin of attraction can neither 
escape nor reach the critical point and must therefore be trapped in the 
boundary surface, until some fluctuation enables them to escape.
For as long as they are trapped, they may follow exotic orbits confined 
to that surface.  

\bigskip

{\bf{Remark:}} 
The best codes for this kind of construction are those that encode only a 
single spin, because multispin codes are generally poor concatenation 
candidates and would therefore correspond to critical points that are so 
unstable that we can ignore them.  This is the reason why we can assume 
that each block generates a single spin without loss of generality.

\subsubsection{The effective Hamiltonian and the RNG flow}\label{hamflow}

The code-tiling intuition also suggests a class of candidate effective 
Hamiltonians for the entangled fixed points.  The system is in the 
ground state of its Hamiltonian by assumption, so we are looking for an 
operator for which these critical point states are eigenstates.  On its own, 
that cannot tell us very much about these Hamiltonians.  However, if the 
system is going to exhibit the desired flow behaviour under renormalization, 
its excited states must have certain properties as well.   

Every error-correction code has a set of error-detection ``check'' operators 
associated with it.  These are a minimal generating set for the stabilizer 
of the code states.  For example, here are the four stabilizer generators 
for the $5$-qubit code we have used as our toy example in this paper:
\begin{center}
\begin{tabular}{|c|ccc ccc ccc|}
\hline 
$\;$ & $1$ & $\;$ & $2$ & $\;$ & $3$ & $\;$ & $4$ & $\;$ & $5$ \\
\hline
\hline
$M_1$ & $\sigma_z$   & $\otimes$    & $\sigma_z$   & $\otimes$  & $\sigma_x$ & 
        $\otimes$    & $\openone$ & $\otimes$    & $\sigma_x$ \\ 
$M_2$ & $\sigma_x$   & $\otimes$    & $\sigma_z$   & $\otimes$  & $\sigma_z$ & 
        $\otimes$    & $\sigma_x$   & $\otimes$    & $\openone$ \\
$M_3$ & $\openone$ & $\otimes$    & $\sigma_x$   & $\otimes$  & $\sigma_z$ & 
        $\otimes$    & $\sigma_z$   & $\otimes$    & $\sigma_x$ \\
$M_4$ & $\sigma_x$   & $\otimes$    & $\openone$ & $\otimes$  & $\sigma_x$ 
        & $\otimes$  & $\sigma_z$   & $\otimes$    & $\sigma_z$ \\
\hline
\end{tabular}
\end{center}
The code space is a pair of common eigenstates of all of these operators, 
with eigenvalue $+1$ (by convention).  These operators can detect single qubit
errors by projection onto the error basis states.  States for which a basis 
error has occurred are also eigenstates of these operators, but at least one 
of the stabilizer generators will return an eigenvalue with the opposite sign 
if such an error has occurred. The stabilizer operators all intercommute, so 
we can simply add the generators, as pointed out by Kitaev \cite{anyonQC}.

Suppose our system has a Hamiltonian $H',$ which is close to the (as yet 
incompletely specified) critical point Hamiltonian, but not identical to it.  
This Hamiltonian can be written as 
\begin{equation}
 H'=H_{\mathcal{C}}+\varepsilon H_{\mathcal{E}},
\end{equation}
where $H_{\mathcal{C}}$ is the critical point Hamiltonian, and that 
$\varepsilon H_{\mathcal{E}}$ is some small perturbation.  Our critical 
point Hamiltonian must have the property that at least some nearby 
Hamiltonians move towards it under the action of the renormalization 
procedure that we had previously obtained for the state. 

If we revert to thinking about our renormalization procedure in active 
terms for a moment, then it is possible to convince oneself that the 
process of active error-correction is equivalent to thinking about 
a perturbation expansion in active terms: instead of just projecting onto 
the eigenbasis of the unperturbed Hamiltonian as a matter of mathematical 
convenience, we instead think of actually performing the corresponding 
projective measurement.  So, the perturbative regime for this Hamiltonian 
will correspond to the regime where the ``correction term'' generates only 
correctable ``errors'', {\emph{provided}} that the eigenbasis of the 
Hamiltonian coincides with that of the check operators for the code.

For example, consider the operator 
\begin{equation}\label{trytileH}
 \mathcal{T} = - K_1 M_1 - K_2 M_2 - K_3 M_3 - K_4 M_4
\end{equation}
where the $K_i$'s are real and positive coupling constants.  (The minus sign 
is needed to ensure that the code space corresponds to the {\emph{ground}} 
states.)  I will call this operator a ``tile-Hamiltonian'', because it only 
includes interactions between spins within the tile.  Therefore it is not
the Hamiltonian for the entire system as it does not include the interactions 
between tiles.  
These $M_i$'s in equation \eqref{trytileH} are a minimal generating set, so 
they do not uniquely define the functional form of the tile-Hamiltonian for 
the $5$-qubit code; any candidate tile-Hamiltonian which contained a complete 
generating set would do just as well.  

We do not yet have an effective Hamiltonian for the system as a whole, because
the tile-Hamiltonian obtained from the stabilizer generators is one which has 
the {\emph{singly}} encoded state as (degenerate) ground state.  The codewords 
of the full concatenated code will always appear as rank $2$ mixtures in any 
finite subsystem consisting of {\emph{intact}} tiles.  

In fact we cannot na\"{\i}vely write down a closed-system Hamiltonian for 
this system.  The usual problems with infinite tensor products will prevent 
us from writing down a global Hamiltonian, even if we didn't mind writing 
down an expression of that size.  We also cannot just look at a subsystem in 
isolation because the dynamics of these systems isn't closed; the system is 
always interacting with larger systems in long-range many-body interactions.  
Such open systems do not have a local {\emph{closed-system}} Hamiltonian, 
they have an evolution superoperator.  
However, we may still be able to construct effective Hamiltonians for these
systems, using techniques from the theory of the dynamics of open quantum 
systems \cite{BrunTraj}.  

There is a result due to Kitaev, in \cite{Kitaev1,anyonQC}, in which he shows 
that if the perturbation term is sufficiently weak and sufficiently more 
localized than the stabilizer generators, then when the state drops back 
into the ground state, it will do so into the ``correct'' ground state 
(critical point) instead of the ``wrong'' one (which would eventually cause 
the system to be overwhelmed by errors and therefore approach some noisy 
sink point).  
In other words, any state that differs from the code state by a correctable 
error will be in some linear combination of the ground (code) states, and some 
excited states.  These eigenstates will themselves be eigenstates of some 
Hamiltonian, which will differ from the code Hamiltonian by a perturbation 
term.  If the ``error'' is correctable, that perturbation must be weak 
(compared with the couplings that generate the code-block behaviour) and it 
must be only weakly nonlocal.  

To obtain the evolution operator for the whole system, we would have to 
concatenate the tile-Hamiltonians and add correction terms to take account 
of the interactions between logical qubits of all orders.  
In order to construct these, we will need the encoded operations for this 
code.  These have the effect of performing $\sigma_x$ and $\sigma_z$ on 
the encoded logical qubit, and are usually denoted $\overline{X}$ and 
$\overline{Z}$ respectively.  
(The bar notation is used to denote encoded qubits, and the Pauli operators 
are often abbreviated to ``$X$'' and ``$Z$'' to minimize notational clutter.)  
For the $5$-qubit code, these are usually chosen to be 
\begin{align}\label{XZstandard}
 \overline{X} &= \sigma_x \otimes \sigma_x \otimes \sigma_x \otimes \sigma_x 
                 \otimes \sigma_x \\
 \overline{Z} &= \sigma_z \otimes \sigma_z \otimes \sigma_z \otimes \sigma_z 
                 \otimes \sigma_z,
\end{align}
but the degeneracy of the concatenated code \cite{PreskillNotes7} means that 
they could also be chosen to be 
\begin{align}\label{XZweight3}
 \overline{X}' &= -\sigma_z \otimes \sigma_x \otimes \sigma_z 
                  \otimes \openone \otimes \openone = M_1M_2\overline{X}\\
 \overline{Z}' &= -\openone \otimes \sigma_z \otimes \openone 
                  \otimes \sigma_x \otimes \sigma_x = M_1M_2M_4\overline{Z}.
\end{align}
The stabilizer generators for the encoded qubits and hence the correction 
terms must be constructed from these, to all orders {\emph{ad infinitum}}, 
with no guarantee that we will ever obtain a closed form for a single block.  
All we can say is that by an extension of the earlier perturbation argument, 
it can be seen that longer-range terms must have correspondingly weaker 
coupling constants.  
That still leaves considerable ambiguity in the effective Hamiltonian, in 
both its functional form and the values of its coupling constants.  This 
ambiguity extends to the tile-Hamiltonian and it can be seen that this 
is a consequence of the fact that the stabilizer for this code is abelian.  
(We will return to this point in section IV.  For now, we'll just add it to 
the list of problems with this example.)

\bigskip

{\bf{Remark:}}
There are two distinct meanings of the phrase ``effective Hamiltonian'' in 
common use.  One term originates in statistical mechanics; this is the 
Hamiltonian of the rescaled system obtained by renormalization.  The other 
meaning of this term is from the theory of open quantum systems, sometimes 
called ``Quantum Operations theory'' \cite{BrunTraj}.   You may well 
encounter one of these effective Hamiltonians if you have a system 
undergoing an open (i.e., locally non-unitary) evolution and you try to 
write down a ``Hamiltonian'' for it.  These operators are typically 
non-Hermitian because they include terms inherited from the system's 
interactions with its environment.   However, there is no particular need 
to change the nomenclature here; it would seem from our investigations that 
``effective Hamiltonians'' in the condensed-matter sense are likely to be 
``effective Hamiltonians'' in the open-systems sense.

\subsection{III.3: Problems with the toy example}\label{toyproblems}

The above example does not really constitute a rigorous argument.
By now, anyone reading this in a critical manner will have noticed a number 
of problems with our toy example.  This is as far as our toy example can 
take us and these problems must now be addressed. 

The first and most obvious problem is that there are a very large number of 
codes and tilings that would seem to work for a given lattice.  Once a given 
tiling has been chosen, the fact that it has a period longer than a single 
lattice spacing means that the choice of position of the tiling is degenerate. 
There is also the fact that for any given code and tiling combination, there 
are typically several inequivalent ways of embedding the code by assigning the 
qubit labels to the various lattice sites in each tile.  This choice can 
affect how non-local the associated tile-Hamiltonian will be.  How do we 
choose which of these many possibilities is the right one for the system 
we are interested in?

The second problem is that the Hamiltonians associated with any particular 
choice of code, tiling and embedding seem to be intrinsically non-local. 
Certainly, this is the case for our $5$-qubit code example, and this may  
well be the case whenever we use a code that is non-degenerate when singly 
encoded.  (Concatenated stabilizer codes are always degenerate, even if the 
original code is non-degenerate \cite{PreskillNotes7}.)  There are 
irreducibly non-local terms in the effective Hamiltonian that do not seem to 
be decomposable into short-range, constant terms.  This is not just an 
{\ae}sthetic problem that makes it difficult to write down the effective 
Hamiltonian in a tidy closed form.  It implies the presence of interactions 
that are {\emph{instrinsically}} long-range, which would seem to be 
unphysical. 

There is also no apparent connection to the native Hamiltonian of the system. 
So far we have started with the state and then written down an associated 
Hamiltonian.  We would like to be able to start with the Hamiltonian and try 
to find out about its ground state.  This is also more natural from the point 
of view of quantum field theory, where we are accustomed to thinking in terms 
of renormalizing the coupling constants in the Hamiltonian, whereas this 
approach just seems to leave these as essentially free parameters.  
There is even some considerable ambiguity about the functional form of the 
Hamiltonian (the ``abelian stabilizer ambiguity''). 

Lastly there is the arbitrary basis change problem, mentioned at the end of 
subsection III.1.  We could argue that the requirement that short-range 
physics washes out under the renormalization group flow would force us to 
choose such an encoding but as this is what we are trying to show, such an
argument would be circular.

\subsection{III.4: Conclusions from this approach}\label{entconc}

So, what can we say for this approach?  There would seem to be good reasons 
to suspect a link between the stabilizer formalism and renormalization. 
The formalism even seems to produce gauge-like symmetries naturally, because 
the choice of basis labelling for the encoded spin is arbitrary, but must 
nevertheless be made; if we were simply to average over all the possibilities, 
the effect would be equivalent to a fatal error syndrome and would destroy 
the very structures we were looking for.  

\bigskip

However, all this is still contingent upon the unproven conjecture that 
systems undergoing quantum phase transitions exhibit entanglement on all 
length scales.  The only plausibility argument for this explicitly invokes a 
pure and non-degenerate ground state, whereas all these code-tiling ansatzen 
have highly degenerate ground states.  These would be locally (and could also  
be globally) mixed at zero temperature.  So, it is still not unreasonable to 
disbelieve the initial assumption on which the entire argument in this 
section is based. 

Even if the entanglement conjecture is correct, there are also many conceptual 
problems with the intuitive picture we have drawn from it, as we have just 
detailed.  
Fortunately, the quantum Kadanoff construction in the previous section (II) 
is logically independent of this conjecture.   In order to resolve these 
problems, we must now return to this foundation and make a shift in our 
perspective once again.

\section{IV: The interaction-based approach}\label{interaction}

We now go right back to the beginning again, to the quantum Kadanoff 
construction.  We will also recall White's insight that we should think in 
terms of the interaction between the block and the surrounding lattice 
but we will make some important modifications.  
Instead of trying to estimate the block-spin and then correct for the 
boundary conditions, we will instead show that the Kadanoff block-spin 
construction can be understood {\emph{entirely}} as a manifestation of the 
interation between the block and the rest of the lattice, up to a (possible) 
gauge-like freedom in labelling the bases for the block-spins.

As it happens, the solution to all the problems in the previous section is a 
very simple one: let the system's dynamics choose the code.  When QECCs were 
first developed they were thought of in terms of active, external 
intervention in the system, designed to drive it in the direction we want it 
to go.  But in this situation, there is no outside intervention by 
assumption, so it would make more sense to look at what the system is already 
doing on its own.

For any readers who are not quantum information theorists, this will mean 
using the theory of decoherence-free subspaces (or, more generally, 
decoherence-free {\emph{subsystems}}).  
This alternative approach to fault-tolerance was invented by different 
groups of people to those who invented QECCs. These constructions were 
originally thought of in different terms, and so were given different names; 
the term ``decoherence-free subspaces'' was first proposed in \cite{DFS1}.  
The reason for the apparent difference is that while QECCs assume that the 
error syndrome consists of independent, single-spin errors, decoherence-free 
subspaces originally assumed a collective error syndrome that affected all 
the spins in the same way.  It was only later that it was realized that the 
two approaches to fault-tolerance could both be understood in terms of the 
stabilizer formalism \cite{DFSconcat} and that decoherence-free subspaces 
were in fact highly degenerate QECCs in the sense of \cite{ThQECCs}.
In between these two extremes we have the theory of decoherence-free 
sub-{\emph{systems}}, which were first introduced in \cite{generalnoise}, 
and their connection to the stabilizer formalism in terms of non-abelian
stabilizer groups was outlined in \cite{Zbridge}. 

We will follow the standard practice of using the acronym ``DFS'' to denote 
both decoherence-free subspaces and subsystems, when no confusion will result. 
Otherwise, we will call these DFS-{\emph{spaces}} and DFS-{\emph{systems,}} 
as appropriate. 
There are a number of very good introductions to the theory of fault-tolerant 
quantum computing using decoherence-free subspaces in the literature (see, 
for example, \cite{KempeDFS}).  For an in-depth introduction into DFS theory 
focussing on their uses as quantum memories (as distinct from fault-tolerant 
quantum computation) and a much more detailed set of references, see 
\cite{DFSSrev}.

It should be noted that DFSs are a property of atypical Hamiltonians: there 
needs to be at least an approximate symmetry in the interaction between the 
system and its environment for them to occur.  This is as it should be, 
as phase-transitions are atypical behaviour.  (This counters the ``arbitrary 
basis change'' objection, above.)  
DFS-spaces are also robust against sufficiently small perturbations 
\cite{DFSrobust}.  Recall that such robustness is an essential feature if 
we are to obtain the RNG flow behaviour characteristic of a critical point.   
(It is also worth noting that the fact that DFS-spaces are highly degenerate 
QECCs would be consistent with our earlier suspicion that one of the problems 
with our toy example was that we had used a non-degenerate code and therefore 
found ourselves dealing with a massively non-local effective Hamiltonian.) 

\begin{theorem}{\bf{(Knill, Laflamme and Viola) 
\newline
[Theorem 5 from \cite{generalnoise}]}}\label{Cstar}
\newline
Let $\mathcal{A}$ be a $\;^{\dagger}$-closed algebra of operators on the 
state space $\mathcal{S},$ including the identity.  Recall that the 
commutant $Z(\mathcal{A})$ is the space of all operators commuting with 
$\mathcal{A}.$  Then $\mathcal{S}$ is 
isomorphic a the direct sum,
\begin{equation}
 \mathcal{S} \sim \sum_i \mathcal{C}_i \otimes \mathcal{Z}_i,
\end{equation}
in such a way that in the representation on the righthand-side, 
\begin{equation}
 \mathcal{A}=\sum_i Mat(\mathcal{C}_i)\otimes I^{(Z_i)}
\end{equation}
and the commutant of $\mathcal{A}$ is given by 
\begin{equation}
 Z(\mathcal{A})=\sum_i I^{(C_i)}\otimes Mat(\mathcal{Z}_i)
\end{equation}
where $Mat(\mathcal{H})$ denotes the set of all linear operators from 
$\mathcal{H}$ to itself. 
\end{theorem}

Necessary and sufficient conditions for a DFS-space to be present in a system 
were found in \cite{DFSconcat} and for a DFS-system in \cite{KempeDFS}.  
These results draw heavily on the representation theory of operator algebras, 
see \cite{RepAlg}.  See also \cite{generalnoise,constructqubit,kribs1,kribs2}.

\subsection{IV.1: The role of the block-lattice interaction}\label{Indra}

In the conventional picture of renormalization, the block-spin is regarded 
as some kind of average property of the block, with the interactions between 
the block and the lattice being accounted for as something separate; as 
boundary effects that force us to adjust the construction as we go along 
\cite{White93algs}. 
By contrast, the usual explanation of renormalization offered in the quantum 
information literature is in terms of the entanglement, as in \cite{ON1}.

The central claim of this paper is that we must incorporate both these 
perspectives, but go beyond them.  We claim that theorem~\ref{Cstar} is 
the algebraic principle underlying the Kadanoff block-spin transformation.  
The justification we are proposing for the Kadanoff construction is that it 
is the interaction between the block and the rest of the lattice which 
defines the block spin in the first place.   This structure can only arise 
when the block-lattice interaction satisfies a suitable symmetry sufficiently
closely to meet the conditions in \cite{DFSrobust}.  For a critical point to 
occur, there must be a hierarchical tiling of the lattice into blocks such 
that the blocks meet this condition.  
For physically reasonable systems where the Hamiltonian consists of 
short-range interactions, these states will typically be highly entangled: 
to hide a spin from local interactions, secrete it in a non-local degree 
of freedom.  Thus we may deduce that their entanglement will have 
the expected fractal structure, with entanglement on all length scales. 

In this picture the block-lattice interaction chooses the tiling, the 
embedding, the code, the block-spin transformation and the equivalent of the 
value of $m,$ the number of states we can keep at each renormalization step.  
There will most likely be little or no freedom to choose these parameters, 
thus answering what we consider to be the most fundamental of Wilson's 
objections to the Kadanoff construction: {\emph{why that particular 
transformation?}}  
The only non-trivial freedom we will have left is in how we choose to label 
the basis states for the block-spin subsystem, which we will interpret as a 
type of gauge symmetry.

\section{V: How might we implement this in practice?}\label{implementation}

We will now outline how we might use these ideas in practice.  For simplicity, 
consider a one-dimensional spin-chain, with your favourite local, periodic 
Hamiltonian.   The procedure will have two main parts.  First of all, we must 
determine our renormalization procedure; we must find the appropriate 
block-spin transformation for this system.  We should then be able to use this 
in the conventional way to determine the critical exponents for the system.

\subsection{V.1: Determining the block-spin transformation}

Begin by choosing a trial block.  (As this is a 1-dimensional system, the 
tiling problem is trivial.)  It is probably best to start with a three-spin 
block.  The smallest-known DFS-space needs four spins \cite{noiselessDFS} 
and the smallest-known DFS-system requires three spins \cite{generalnoise}.  
Look for a DFS in this system that encodes a single spin. (Multispin DFSs 
will have the same concatenation problems as multispin QECCs.)  

There are at least two potential difficulties here.  First of all, the 
conditions for an {\emph{exact}} DFS are only satisfied on lower-dimensional 
sets in parameter space.  Fortunately, these structures are very robust 
against perturbation \cite{DFS1,DFSrobust}. 
That still leaves the problem of how we should determine whether we are in
``sufficiently close'' to satisfying the DFS conditions to be in the 
stability regime for a DFS.  Unlike conventional renormalization, we 
don't start out knowing what our renormalization procedure should be: we 
have to find it.  For the time being at least, we will need to know 
approximately where the critical point is located in parameter space.   

This statement immediately implies that we will need to devise some form of 
search procedure.  Such a search procedure implicitly relies on the existence  
of some kind of distance measure.   How close would we need to get to the 
basin of attraction of a critical point before we could detect its presence?  
The structure of this basin of attraction suggests that this should be 
possible.  The framework in \cite{Zbridge} implies that we should be able to 
apply Kitaev's result from \cite{Kitaev1,anyonQC} to this problem.  Together 
they imply that the basin of attraction will have a layered structure.   
This is because both conventional QECCs and DFSs can correct errors in two 
different ways.  The first way is for trivial errors that consist of 
operations in the stabilizer.  The second is the active correction performed 
by recovery operators \cite{Zbridge,DFSSrev}.  Conventional QECCs rely 
primarily on active correction, but can be thought of as DFSs in their 
passive regime.  DFSs rely primarily on passive correction, but one can still 
define an active recover operator for these if required. 

The innermost layer is where the DFS behaves as an ``error-avoiding'' code: 
the errors (differences in the Hamiltonian from the true fixed point) are 
trivial because they are in the stabilizer of the state and so have no 
effect. Then there is the ``active correction'' regime, where the DFS can be 
understood as behaving as an active QECC, so we can still define a recovery 
operation and hence a renormalization procedure.  
This single-encoding error structure could then be concatenated, as per 
Kitaev's theorem.  Error terms that might be fatal for a single encoding may 
end up in the stabilizer of the concatenated code or the active correction 
regime, and so on.  The renormalization would still be able to find the 
fixed point, if we only knew what the renormalization was.  

So, we are still left with something of a chicken-and-egg problem.  This is 
one of the reasons why this outline does not constitute a precise description 
of how to implement this procedure.   For the time being at least, we would 
need to know where the critical point is located in order to be able to use 
this procedure at all.  The problem of determining how close is close enough 
will have to await further work; we anticipate that it will require a 
generalization of the results in 
\cite{Kitaev1,anyonQC,DFS1,DFSrobust,generalnoise}.

\bigskip

The second problem is caused by the fact that the existence proofs for DFSs 
in \cite{DFSconcat,KempeDFS} depend on the Kraus Representation Theorem 
(see \cite{Krausrep}, or alternatively, \cite{PreskillNotes3,Mike-n-Ike}).  
This result is known to break down for the case of open system evolutions 
when there is prior entanglement between the system and its environment 
\cite{Pechukas1}.  There is also no known way to make the Kraus representation 
theorem fully Lorentz covariant \cite{PeresTernoRev}.  
In a realistic system, there will typically be entanglement between the block 
and the rest of the lattice, indeed if we believe the conjecture that systems 
undergoing quantum phase transitions exhibit entanglement on all length 
scales, we should expect this. 

A quick and dirty solution to this problem is simply to ignore this 
prior entanglement, proceed assuming that the Kraus representation 
theorem is still applicable, and treat any effects caused by the entanglement 
between the block and the rest of the lattice as correction terms.  

Fortunately, a much more sophisticated approach is possible that bypasses the 
difficulties the Kraus representation theorem has with block-lattice 
entanglement.  In fact, this was implicit in \cite{generalnoise}.  Working 
in terms of the error operators directly instead of the states removes the 
need to trace out the bath, and thus avoids the need to assume that the 
system and the bath are initially decoupled \cite{DLidar}.  These ideas are 
developed further in \cite{ObsInd1}.   Indeed, there is an explicit algorithm 
for computing the structure of the interaction algebra in \cite{kribs1}. 

\bigskip

Suppose we fail to find a DFS at the first attempt.   If so, the next step 
should be to abandon this trial block and try again with a new, larger, block 
by including another spin.  (If we were working with a higher-dimensional 
system we should also be prepared to search over different concatenated 
tilings.)
If we started this procedure sufficiently close 
to a critical point, then this search method should eventually find a DFS.   
In fact, DFSs occur in families in systems which exhibit them, but in 
practice only one DFS can be used, as there is a loss of phase coherence 
between the different DFSs in the system.  
We can invoke the arguments in \cite{Zbridge} and define the renormalization 
procedure to be the data-recovery superoperator as before, 
but in order to define that, we must choose which subspace (or subsystem) to 
be the codespace.   With DFS-{\emph{spaces}}, this is generally unproblematic; 
there is a natural, obvious choice.  
With DFS-{\emph{systems}} however, this may not be the case; sometimes there 
is more than one reasonable candidate.  If so, we might need to repeat the 
next step a few times with different choices for the encoded block-spin. 

We now keep our newly decoded block-spin, and trace out the decoherence-full 
subsystems.   This is the block-spin transformation needed for the Kadanoff 
construction.   Note that  the transformation is unique, up to a possible 
choice of basis-labelling of the block-spin (and apart from some mathematical 
ambiguities in codespace choice that should be resolved in time with 
further work).  Note that we have made almost {\emph{no choices}} here; the 
block-spin transformation is entirely dictated by the dynamics of the system. 

We can now invoke the periodicity of the system dynamics to rescale the entire 
lattice in the same way; any DFS structures we found should be congruent with 
the symmetry properties of the whole-system dynamics. 
This decode and discard step completes the first iteration and leaves us 
with a rescaled state and lattice. We can then proceed to the next step.

Suppose for a moment that we had found our DFS by assuming there was no 
inital entanglement present, and using the test in \cite{DFSconcat} or 
\cite{KempeDFS}.  
The correction terms inherited from our assumption that there was no prior 
entanglement between the block and the rest of the lattice would then become 
the rescaled interaction terms.  A similar rescaling should also be produced 
by the approach in \cite{generalnoise,ObsInd1}, although this is rather less 
obvious in the machinery of this construction. 
Note that we have also obtained a rescaled effective Hamiltonian by this 
procedure, complete with rescaled coupling constants.  This is much more 
informative than the approach described in section III where these were left 
as free parameters.

We now search for a new DFS in the rescaled system.  At each step we need to 
record the DFS, the effective Hamiltonian and the corresponding recovery 
operation.  In the original formulation of the Kadanoff construction, the 
system's dynamics are exactly scale invariant and the fixed points of the 
transformation were assumed to be strictly pointlike.  If we insist on that 
property, then there is no obvious guarantee that the DFS from the next 
iteration will have the same form, but other limit cycles are possible 
besides pointlike ones.

\subsection{V.2: Moving away from the fixed point}

If we are only interested in the structure in the immediate vicinity of the 
fixed point, then we would be done.  If we want to use our 
renormalization procedure in a more conventional way, we need to move on 
to the next stage. 
The conditions for the existence of a DFS are only satisfied for atypical 
Hamiltonians that exhibit a particular symmetry property.  How can we 
apply this construction to the generic case, which occurs precisely 
when these conditions are not satisfied?   Recall that Wilson's 
renormalization method tacitly assumed a block-spin structure in the vicinity 
of the critical point, as well as at the fixed point itself.  Under these
conditions, the iteration flows away from the critical fixed point, into an 
attractor or sink point.  Now that we have found the concatenated DFS 
for the critical point, we can invoke the unified stabilizer picture in 
\cite{Zbridge} again and consider the renormalization as an 
explicit code recovery super-operator.  
The point of doing this is that we can always implement an explicit recovery 
super-operator on a state of the right size, even if that state is 
{\emph{not}} within the stability region surrounding a concatenated-DFS 
critical point.  
We can still perform the recovery operator concatenation, and it will 
converge on {\emph{something,}} but not the critical point.  If we attempt 
to implement a concatenated coding scheme on a system that is outside the 
fault-tolerance threshold for that scheme, further layers of encoding will 
make the state more noisy, not less so.  In terms of a quantum computer, 
this will correspond to a catastrophic failure of the code and the notional 
``computer'' will undergo a runaway error syndrome, thus moving the system 
further and further away from the critical point, as required.  This 
superoperator concatenation can be used to renormalize any observables in 
the obvious way.

\subsection{V.3: A few remarks}

This is not a complete specification by any means, but it could provide  
a useful framework for future investigations \cite{follow}.  A fully worked 
out algorithm would need to contain a proper search routine, and that 
will require the development of an appropriate distance measure. 
This might need to be the entanglement fidelity \cite{Entfidef} as 
this is the appropriate measure for quantifying the effectiveness of a 
quantum error correction scheme \cite{EntfidN}.   Having said that, provided 
the code performs sufficiently well for all pure states, the simple fidelity 
will do \cite{ThQECCs}.  Likewise, a fully Lorentz-covariant version of 
the Kraus Representation Theorem would greatly improve this method. 

\bigskip

There seems to be no particular reason to assume that only one DFS 
concatenation will be found.  If these occur at slightly different 
positions in the parameter space, these would constitute a fine 
structure to the renormalization group flow.  If they happen at the 
same values of the system parameters, they should reflect some kind 
of equivalence class structure among concatenated DFSs; either that, or 
the would seem to be some ambiguity in the RNG flow defined by this 
procedure. 

\bigskip

The procedure outlined above employs a concatenated tiling renormalization. 
A more conventional ``progressive addition'' method would add a single new 
spin to the original block at each step, and the next step in the 
renormalization uses essentially the same block.  The standard DMRG is an 
example of a progressive addition renormalization.  
We believe that this modification may not be benign, in the sense that they 
may not be equivalent to a concatenated tiling renormalization and so should 
not be used interchangeably.   However, that is not to say that such addition 
renormalization schemes are incorrect, but rather that they correspond to a 
different encoding structure, which we discuss further in the Appendix.  

\bigskip

Lastly, I would like to highlight an important qualitative difference between 
this renormalization scheme and the toy example in section III.  One of the 
objections to this example made in subsection III.3 is that it seems to 
require irreducibly non-local interactions in its associated Hamiltonian 
and that these are unphysical.  Contrast this with the way we obtained our 
rescaled Hamitonian in the procedure we have just outlined above.  Here the 
non-local terms arise entirely as residual correction terms from the blocking; 
the native Hamiltonian is actually local.  Thus the {\emph{effectively 
non-local}} interactions that emerge under iteration of this renormalization 
are entirely {\emph{mediated}} by the discarded subsystems we have chosen to 
ignore in our renormalization.  No {\emph{intrinsically}} long-range 
interactions are required in the native Hamiltonian at all.

\subsection{V.4: Relationship to the DMRG}\label{relDMRG}

Although the thinking behind this renormalization procedure looks rather  
different from that behind the DMRG the previous methodology is consistent 
with it, where it works.  Those situations where it {\emph{does}} seem to 
make sense to think of the block-spin as an average of the physical spins 
can be understood in the coding picture; in this case, the code is what is 
known as a ``na{\"{\i}}ve repetition'' code, and the decoding operation is 
simply to take the majority vote.  Thinking in terms of a code also makes 
the otherwise rather artificial rescaling of the spin variable seem more 
natural as well.  (It is interesting to note that the direct quantum 
analogues of na{\"{\i}}ve repetition codes do not exist; they are forbidden 
by the No-Cloning Theorem \cite{NoCloning}.)

In fact the connection can be made even closer than that, as we can 
incorporate excited states as well.  Note that in the $5$-qubit code example 
in section III, the recovery operation is one of the transformations that will 
exactly locally diagonalize this state, and that in order for the recovery 
operator to work, the excited states of the system must also satisfy certain 
properties.  
A caveat is needed at this point: as the ground state is degenerate 
in this example, the exact diagonalization transformation is non-unique. 
Note that if we just chose one of the diagonalizing transformations at random 
each time, the effect of this would be to induce a random error in each  
encoded qubit on decoding, which is an error syndrome this coding scheme 
cannot withstand; the RNG flow would therefore not converge on the critical 
point but rather on some noisy sink point.  

The relationship between the two approaches will not always be so close.  In 
this paper we have argued that the principle underlying renormalization can 
be understood as a hidden, non-local structure in the local density matrix 
and that the goal of a renormalization step is to find that structure and 
pick out some factors in that tensor product, while discarding others.  
Compare this with the way the DMRG can be understood in terms of the ordered 
Schmidt decomposition \cite{ON1}.  It is already known in quantum information 
theory that strange things can happen when the ordered Schmidt decomposition 
is applied to tensor products, albeit from a completely different context 
\cite{catalysis1}.  

Sometimes the sorting operation preserves the tensor product structure, 
sometimes it doesn't; it depends on relative sizes of the gaps between the 
eigenvalues in the various subsystems.  So for example, if one subsystem was 
degenerate, with its eigenvalues taking the almost same values, while the 
rest of the system had a strongly non-degenerate eigenspectrum, then 
provided the system with the nearly degenerate eigenspectrum was the 
``correct'' one to keep, then the DMRG should pick it out correctly.  Systems 
like this should exhibit strongly stepped eigenspectra.   But if the sizes of 
the gaps between the eigenvalues are roughly comparable, things can get very 
messy indeed.  In general, there is no simple way to reconstruct the tensor 
product structure of a combined system from its ordered Schmidt 
decomposition alone and this is likely to be a particular problem with 
highly degenerate systems.

\section{VI: Entanglement, Entschmanglement...}\label{altcorr}

The concept of the order parameter for a phase transition and the various 
spin-spin correlation functions have proved to be enormously powerful 
tools in statistical mechanics.  Unfortunately, the spin-spin correlation 
function  doesn't work for quantum phase transitions \cite{Dorit,anyonQC} 
and the order parameters for these systems are notoriously elusive.   
We would like to have equally useful equivalents for quantum phase 
transitions, and if we can relate these to a renormalization procedure, so 
much the better.  
Various possibilities have been suggested as candidates for these, such as 
the entanglement length in \cite{Dorit}, but this will be problematic 
in the infinite lattice limit, as the bipartite separability problem is 
NP-{\small{HARD}} \cite{Gurvitsep}.  It is also not clear how to relate it 
to the renormalization procedure described in this paper.

\subsection{VI.1: The trouble with entanglement measures}\label{trouble}

A number of authors have suggested using measures of multipartite 
entanglement for mixed states to try to understand these systems, in direct 
analogy to the spin-spin correlation function.   This approach has been 
rather hamstrung by the lack of closed forms for these quantities, but even 
if we did have a full set of closed forms for multipartite entanglement 
measures, this approach would still have some serious problems.  
This is because it can be shown via a simple dimension-counting argument 
that the number of algebraically independent multipartite entanglement 
measures grows exponentially in the number of parties 
\cite{mschmidt,2ways} (assuming no one party controls a majority share).  
A similar argument can be used to show this number is doubly exponential in 
the number of parties for mixed states \cite{mixedcount}, although in this 
case that figure includes classical correlations.  

These different entanglement types are not just algebraically independent, 
they are physically inequivalent as well.   For example, in the $5$-qubit 
example used in section III of this paper has no two or three qubit 
entanglement present whatsoever; the concurrence \cite{concurrence} for any 
two qubits drawn from this state is identically zero.  It can also be seen 
that there is no three-qubit entanglement present, as any $3$-qubit reduced 
density matrix is proportional to the identity.  In fact this example is an 
extreme case because the single encoding is non-degenerate and so coding 
existence bounds require that subsystems below a minimum size have density 
matrices proportional to the identity \cite{PreskillNotes7}.  
Some of these bounds may not apply to DFSs, as they are highly degenerate 
codes.   

Nevertheless, it would be unwise to assume that any one entanglement measure 
will be sufficient to detect the entanglement present in any given phase 
transition.  This is because they have enlarged stabilizers, which has 
implications for their entanglement properties:    

\begin{theorem}{\bf{(Carteret and Sudbery)
\newline
[Theorem 2 in \cite{3qubits}, generalized for mixed states]}}\label{elastic}
\newline
Let $\mathcal{H}$ be the space of states, and let $G$ be the group of 
local unitary transformations of $\mathcal{H},$ and let the dimension of 
the stabilizer of a {\emph{typical}} state be zero (which it is, for mixed 
states, \cite{mixedcount}).  Let $I_1,\ldots I_k$ be a set of $k$ polynomial 
invariants which generate the algebra of local invariants in a neighbourhood 
of a state $\rho.$  If the stabilizer of $\rho$ in $G$ has non-zero dimension, 
then there is a linear combination of $I_1\ldots I_k$ which has a stationary 
value at $\rho.$
\end{theorem}

This theorem certainly applies to the abelian stabilizer groups of 
conventional QECCs and it should also be true of the stabilizers of 
DFS-systems \cite{Zbridge} if the non-local error syndromes these correct 
correspond to a simple merging of parties.  (E.g., they can correct non-local 
error syndromes that act on particles 1 and 2 jointly, or on 3 and 4 jointly, 
but not on 2 and 3, say.)

Theorem~\ref{elastic} does not simply that states with enlarged stabilizers 
have maximal entanglement; they could also have minimal entanglement (such as 
the maximally mixed state) or, more typically, a combination of both maximal 
and minimal entanglement, or even saddle-points in the manifold of non-local 
invariants.  
The moral of the story is that states with enlarged stabilizers have highly 
specific entanglement properties.  Note that the converse is not necessarily 
true; an enlarged stabilizer is not a necessary condition for extremal 
entanglement properties.  The fact that the theorem is one way only may mean 
that DFS-systems whose stabilizers do not correspond to a simple merging of 
parties may also have extremal entanglement properties, but any proof of this 
will need a more sophisticated argument than that in theorem~\ref{elastic}.

The practical implications of this are that if we were going to use these 
entanglement measures as our initial probe of a system that we know nothing 
else about, we would have to calculate almost all of these functions, even 
though there are exponentially many.  
If we omitted to check too many types of entanglement, we would risk missing 
the very structures we were looking for.
To make matters worse, for each $n$-party measure, for which $n$ particles do 
you calculate it {\emph{for}}?  Even if we assume that it only makes sense to 
evaluate these functions for particles in contiguous clusters, that will still 
leave us with so many choices that this method would be impractical, even if 
we had a complete set of closed forms for multipartite entanglement measures, 
and could calculate each individual function with ease.

\subsection{VI.2: Order parameters}\label{orderp}

Although this paper has concentrated on the critical fixed points of the 
renormalization group flow, at least some ideas from this picture can be 
applied to the completely repulsive fixed points mentioned in section II. 
The DFS renormalization picture suggests a very natural choice for an order 
parameter for these systems.  This is the block-spin transformation symmetry 
we have invoked to define the entire theory and which is in turn directly 
linked to dynamical symmetries of the system. 
Recall that in section II we supposed the block-spin transformation was an  
exact symmetry of the system, but that as we moved away from the critical 
point, the symmetry would eventually break down, as in subsection V.2.  In 
the infinite lattice limit there will be a sharp transition between the two 
basins of attraction, which produces the necessary singularity in the system's 
behaviour.

It now remains to find a convenient way to quantify this symmetry, preferably 
as part and parcel of the renormalization procedure. 
As we have shown a close relationship between the theory of fault-tolerant 
quantum memory and quantum phase transitions via our quantum analogue of the 
Kadanoff construction, it is natural to look to the theory of fault-tolerant  
memory thresholds for inspiration.  
If we once again treat small changes to the system Hamiltonian as if they 
were noise syndromes, we can use some of these results.  In 
\cite{Approxconcat,Exactconcat} a technique was developed for calculating 
exact quantum memory thresholds for concatenated codes, using the theory 
of quantum channel capacites (see also \cite{Shorchannelrev}).  
Given the error syndrome (correction to the Hamiltonian) and the encoding 
(which we will have found as our RNG) the net effect of the error and 
recovery superoperators can be represented as a quantum channel.  One of the 
examples they calculate in detail is the case the $5$-qubit under a simple 
error model.  

In finite systems the quantum capacity of this channel can take a continuous 
range of values between 0 and 1 as the fault-tolerant quantum memory 
threshold is crossed, but in the infinite concatenation (i.e., infinite 
lattice) limit, the quantum channel capacity of the encoding must be either 
1 or 0.  This quantum channel capacity therefore has precisely the right 
properties to be an order parameter.
Note that while there are many ways to define channels and hence channel 
capacities in these systems in terms of various fictional or actual 
measurements, we claim that the channel defined by the renormalization 
iteration is the only relevant one.

\subsection{VI.3: A natural correlation functional}\label{functional}

This does not yet give us an analogue of the correlation length.  To see 
what this should be, note that for the case of a concatenated code that 
only just fails to withstand an error syndrome, each successive layer of 
concatenation actually makes the problem worse, not better.   So, for such 
a failing code, how many layers of concatenation can we have before the 
quantum channel capacity falls away to less than some small threshold 
$\varepsilon < 1$?  Note that it wouldn't make much sense to specify a 
capacity of less than 1 here, as if the encoding achieves a capacity of 1 
for finitely many layers, it will achieve 1 for infinitely many and we must 
be at the critical point.  We would like something that behaves in an 
analogous way to the correlation length.  It should therefore be able to 
take a continuous range of values between 0 and infinity, so we should 
specify an $\varepsilon < 1,$ strictly (though we may well find that we 
can take $\varepsilon \to 0$ in many cases).  

Therefore it would make sense for us to define the 
{\bf{quantum $\boldsymbol{\varepsilon}$-memory support}} 
to be the maximum size of a region of the lattice that is able to support 
such an emergent spin, by analogy to Kadanoff's definition of the 
correlation length.   
Suppose that it takes $r$ layers of encoding for the quantum channel capacity 
to fall below $\varepsilon,$ and each Kadanoff block contains $k$ spins with 
a lattice spacing of $L.$  If our system is one-dimensional, then the quantum 
$\varepsilon$ memory support will be $k^rL.$  Likewise for a two-dimensional 
system with $L$ spins in each block, we will obtain an $\varepsilon$-memory 
support of $k^rL^2$, but this time with units of the lattice-spacing squared.  

This function will typically scale with the dimension of the lattice, unless 
we choose a really strange tiling which is effectively lower-dimensional.
Naturally the precise numerical value of $r$ will depend on our choice of 
$\varepsilon,$ but its scaling characteristics for fixed $\varepsilon$ 
should not.   In particular, this function will be infinite if we are 
sufficiently close to a critical point, as we will be within the quantum 
memory threshold for the DFS concatenation.  

This is not (strictly speaking) an entanglement measure, but it will be 
closely related to the entanglement properties for any realistic system, 
because DFSs for these systems will be entangled states.  
This will also ensure that there is a close link to the yield of the 
corresponding (finite-batch) entanglement distillation protocol, 
particularly if we believe that the convergence of the renormalization 
procedure should be defined with respect to the entanglement fidelity in 
\cite{Entfidef,EntfidN}.

\section{VII: Towards a continuum limit}\label{continuum}

We have still not managed to express this renormalization procedure in 
differential form (c.f. \cite{Wilson1}) although the results in 
\cite{Approxconcat,Exactconcat} are a step in this direction.  
At the moment, we have an iteration that we would like to convert into a 
system of differential equations that we can solve exactly.
This is not as easy to do rigorously as it may first appear.  To see why 
this is non-trivial, consider the opposite problem, where we have a set of 
differential equations that we would like to approximate by an iteration, 
for numerical analysis. 

It is well known that simply replacing the derivative with the corresponding 
finite-difference doesn't always work, as the flow trajectories may be highly 
sensitive to small changes in the initial conditions.  The fixed points of 
the system may also be affected.  An example of this problem can be seen in 
the logistic equation.  The continuous (derivative) version of this is 
\begin{equation}
 \frac{dN}{dt}= r \left(1-\frac{N}{K}\right)N
\end{equation}
where $N$ is the population, and $r$ and $K$ are parameters.  The 
corresponding finite difference equation is
\begin{equation}
  N_{n+1} = \mu \left(1-\frac{N_n}{\kappa}\right)N_n,
\end{equation}
obtained by substituting $dN/dt = (N_{n+1}-N_n)/\Delta t$ and then writing  
$\mu = 1+ r\Delta t$ and $\kappa = (1+r\Delta t)K/(r\Delta t).$  These 
two equations share two fixed points (the extinction and steady-state points) 
but while the finite-difference equation exhibits period-doubling and even 
fully chaotic behaviour for sufficiently large values of $\mu,$ the 
corresponding differential equation does not.  This is admittedly a very 
simple example, but chaotic flows have been reported in some numerical 
renormalization schemes \cite{Binney4}.  
Note that the finite difference equation is not a single equation, but rather 
a family of equations whose parameters $\mu$ and $\kappa$ are functions of 
our choice of stepsize, $\Delta t$ and thus the point at which the onset of 
chaos occurs will be a function of the stepsize, if plotted as a function 
of the original system parameters $r$ and $K.$  

There are some results in numerical analysis that tell us when an iteration 
is a reliable approximation to a differential equation.  These results are 
known as ``Shadowing Theorems'', and they all seem to rely on being able to 
take the step size to zero.   We have the opposite problem in that in our 
case, the iteration is the ``real'' system and the differential equation is 
the approximation.  Thus our stepsize is fixed at $1$ and we are not free to 
take it to zero.   Having said that, there clearly are situations where this 
approximation is valid.  A closer examination of the critical points involved 
in these cases reveals that they are insensitive to small changes in the 
initial conditions and indeed, they typically feature merging trajectories; 
multiple different starting points end up at the same fixed point and thus 
the system can be said to forget its initial conditions.  In other words, 
the system's flow is irreversible. 

There has been considerable recent interest in the question of whether or 
not renormalization group flows are reversible or not 
\cite{PreskillFD,finegloss}.  The problem above suggests a slightly different 
perspective on this question: it might just be that irreversibility is a 
necessary condition for a continuum limit to be trustworthy.  Without some  
applicable converses to the shadowing theorems it is impossible to be more 
specific than that; this question will also have to await further work. 

\bigskip

So, does our reinterpretation of the Kadanoff picture give rise to an 
irreversible RNG flow?  This question can be addressed through the 
thermodynamics of quantum error-correction \cite{thermocorrect}.  This is 
because the error-correction process produces classical information about 
the error syndrome as a by-product.  Once we have undone the error, the 
results of the syndrome measurement must be erased.  Landauer's principle 
states that erasure has a thermodynamic cost of $k\log(2)$ per bit.
We can minimise this cost by using efficient reversible data compression, but 
we cannot eliminate it.
This result goes through for the case of the ``passive'' error correction 
of DFS theory, as the error syndrome will be discarded with the 
decoherence-full subsystem.  
It would seem likely that this condition will be met for the critical points 
described in this paper.  Whether or not it would be met by the completely  
repulsive fixed points will most likely depend on whether we can find 
fixed-step-size converses to the shadowing theorems.

\section{VIII: Conclusions and further work}\label{conc}

In this paper we have tried to understand renormalization theory in terms 
of quantum information theory.  In the process we found ourselves going right 
back to the Kadanoff picture in order to build a bridge between these  
two conceptual systems.  The Kadanoff picture has traditionally been regarded 
as not much more than a ``Just So'' story, with only intuitive value that 
cannot be made rigorous.  We believe this evaluation should be reconsidered.  
We then incorporated White's insights into the role of the interaction between 
the block and the rest of the lattice {\emph{a fortiori}}.  This led us to 
conclude that the Kadanoff block-spin structure can be explained as a 
consequence of the block's interaction with the rest of the lattice.  
The block-spin is created by the rest of the lattice.  Every pearl in Indra's 
net not only reflects every other, it {\emph{consists entirely}} of those 
reflections. 

Strangely, despite the fact that the conceptual development of this paper 
began with a conjecture about the entanglement properties of these systems, 
we have ended up concluding that entanglement measures may give little 
insight into these systems if used in {\emph{isolation}}.  Rather we claim 
that the key to understanding these systems lies in realizing that the 
Kadanoff block-spin is induced by the interaction between the block and the 
rest of the lattice.  It should be possible to interlace the entanglement 
properties with the renormalization procedure in a natural way, thus 
mitigating the problem in theorem~\ref{elastic} to some extent and enabling 
us to probe these in greater detail.  
I believe the papers \cite{NatEnt,EntObs,Ebeyond} contain important steps in 
this direction, even though they do not mention renormalization 
{\emph{per se}}. 

The symmetry properties of the system's dynamics feature prominently in our 
model of renormalization.  The condition for the existence of such a DFS 
structure is a symmetry condition and thus leads naturally to an order 
parameter.  We can also define an analogue of the correlation length.
While we have not proved that quantum phase transitions must necessarily 
exhibit entanglement on all length scales, we have shown that this is 
to be expected under the physically reasonable conditions of short-range 
interactions. 

We are still quite some way from giving an explicit alternative to existing 
numerical renormalization procedures.  There are many unresolved problems 
in the detection of DFS structures, the generalization of the Kraus 
representation theorem, and possible converses to the Shadowing theorems.  
We hope that the details required will become clearer in further work 
\cite{follow}.  

There is also the intriguing possibility of some kind of gauge symmetry in 
the choice of basis-labelling for the encoded qubits.  It is not yet clear 
to me whether this freedom is congruent to known gauge symmetries or not.  
(Note that the choice of gauge may need to be time-dependent; some DFSs 
exhibit Lamb shifts \cite{LambDFS}.)  

The reformulation of renormalization theory given in this paper is entirely 
in terms of local density matrices and open-system dynamics.  As such, it 
should be straightforward to apply this to finite-temperature systems.  This 
is because the mathematics we have used sees errors only as undesired 
admixtures; it {\emph{does not care}} whether these are caused by unwanted 
entanglement with an external environment, ``proper'' thermal mixing, or 
any combination of the two.  Therefore it should be possible to extend this 
framework to incorporate quantum-to-classical transitions, as these will be 
the completely repulsive fixed points (or rather, surfaces) introduced in 
section II. 
However, we anticipate that the problems with converse shadowing theorems 
will be much more severe for these cases, not least because they are unlikely 
to be pointlike structures in the parameter spaces. 

\bigskip


\begin{acknowledgments}

\noindent
{\bf{Acknowledgments}}

Over the course of the 18 months or so that it has taken me to write this 
paper I have enjoyed some very interesting discussions with the following 
people: 
Michael Berry, Robin Blume-Kohout, Louis Crane, Viv Kendon, 
Raymond Laflamme, Fotini Markopoulou-Kalamara, Michele Mosca and David Poulin. 
I would also like to thank Daniel Gottesman and Guifr{\'{e}} Vidal
for asking some wonderfully difficult questions, and especially 
Todd Brun and Lee Smolin for also patiently reading numerous versions of this 
manuscript and their invaluable constructive criticism of those drafts. 
\newline
\newline
I am also grateful to Daniel Lidar for drawing to my attention a subtlety in 
reference \cite{generalnoise} (and also \cite{ObsInd1}) which should provide 
a rather elegant way to fix one of the problems with the renormalization 
procedure I proposed in version 1 of this paper. 
\newline
\newline
This work was supported by the IQC at the University of Waterloo, the LITQ in 
DIRO at the Universit\'e de Montr\'eal, MITACS, The Fields Institute and the 
NSERC CRO project ``Quantum Information and Algorithms''.

\end{acknowledgments}


\section{Appendix: Some other examples}\label{examples}

Here we discuss some other examples which may be of interest.  They also 
serve to further illustrate some points which were only mentioned in 
passing in the main text.

\bigskip

\noindent
{\bf{Some translationally invariant solutions}}

\begin{figure}[floatfix]
    \begin{minipage}{\columnwidth}
    \begin{center}
        \resizebox{0.5\columnwidth}{!}{\includegraphics{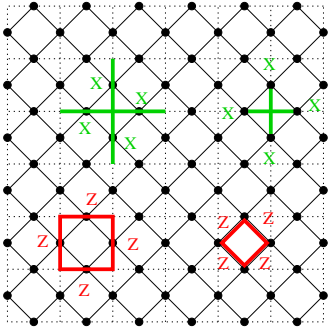}}
    \end{center}
    \end{minipage}
    \caption{Minimal generators for a toric code.  Top left: an $X$ ``site'' 
    generator drawn on the dual lattice.  Top right: the same generator 
    represented on the physical lattice.  Bottom left: a $Z$ ``plaquette''
    generator on the dual lattice.   Bottom right: $Z$-plaquette on the 
    physical lattice.}
    \label{toricdef}
\end{figure}

Another class of solutions which are also a type of stabilizer code are 
the toric codes \cite{anyonQC}.  To show this, we need to establish that 
these codes admit a recursive renormalization scheme.
Toric codes are usually defined on a dual lattice, where the spins live on 
the edges rather than the vertices.  Figure~\ref{toricdef} illustrates the 
generators for these codes, on both the dual and physical lattices.  

\begin{figure}[floatfix]
    \begin{minipage}{\columnwidth}
    \begin{center}
        \resizebox{0.8\columnwidth}{!}{\includegraphics{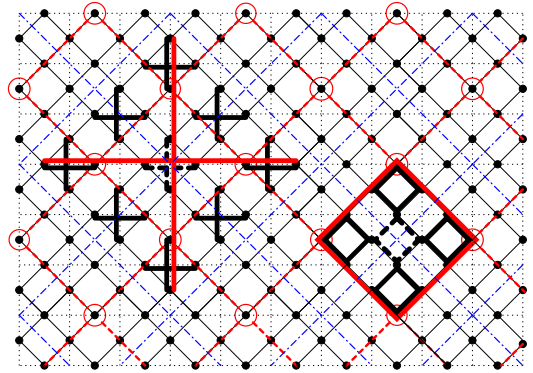}}
    \end{center}
    \end{minipage}
    \caption{Rescaled generators for a toric code.  The black generators are 
    the standard ones, and those drawn in grey (red) are from an alternative 
    (minimal) generating set.  These longer-range generators can be 
    constructed from the short-range generators of the same type 
    ($X$-site or $Z$-plaquette) and the short-range generators shown in 
    dashed lines can then be omitted.  The block spins are denoted by open 
    (red) rings, and the rescaled lattice is drawn in thicker, dashed (red) 
    lines.  The unit cell for this rescaling is shown with  dot-dashed 
    (blue) lines.}\label{toricsoln}
\end{figure}

Unlike the $5$-qubit example in section III, these solutions are 
degenerate codes, and they are translation invariant.  They satisfy the 
Kadanoff construction, because it is possible to choose a minimal generating 
set that is scale invariant, using a construction which is illustrated in 
figure~\ref{toricsoln} \cite{nestedtoric}.  (This seems to be related to 
the ``expanding diamonds'' construction \cite{expandingdiamonds}.)

The stabilizer generators for these codes are usually chosen to have 
interactions between no more than four spins each.  
Minimal generating sets are not unique, so we are free to choose a different 
one with a more obviously hierarchical structure.  (Here the abelian 
Hamiltonian ambiguity is useful, rather than being a problem.)
The large $X$-site generator is the product of the $9$ smaller ones 
(one is hidden under the middle of the larger generator).
If a {\emph{minimal}} generating set is desired, we can then omit one 
of the smaller generators and replace with the large one.   Likewise, the 
larger $Z$-plaquette generator is the product of the $5$ smaller ones it 
contains, and we can choose to omit the small plaquette in the middle, as 
the new, longer-range plaquette has made it redundant.
Note that there is more than one way of doing this.  The construction 
can be displaced by integer multiples of the lattice spacing, and so is 
translation invariant.  We could also choose to make the rescaled generators 
larger, so each cycle would scaled by a factor larger than $3.$   
Thus the corresponding effective Hamiltonians will have the required 
self-similar structure. 

Note that every spin is in only one unit cell.  The tiling does not overlap, 
but the tiles do interact with each other directly, as they share some 
stabilizer generators. 
The long-range interactions in these systems can claim to be genuinely 
emergent, because all the longer-range interactions arise as products of 
four-spin interactions.  
The corresponding renormalization procedure is a progressive, 
fault-tolerant decoding operation.  This is a special case of the 
procedure is described in \cite{topomemory}.  
We can either modify the triangulation of the surface one move at a time, 
or choose to collect the moves into batches that follow our choice of 
blocking for the lattice.  

\bigskip

Toric codes certainly possess short-range entanglement and they must have 
global entanglement in order to function as error-correcting codes.
However, proving that they have entanglement on all length scales requires 
that we take some care with our definitions. 

Suppose we try to ascertain whether these codes have n-party entanglement 
by looking for blocks of $n$ spins, and checking to see if they are 
entangled.  If we find a block that is, we then look at blocks of size $n+1$ 
that include our entangled block of $n$ spins and repeat. 
This calculation was done a few years ago by D. Aharonov and Gottesman 
\cite{ToricEnt}, using a reasonable measure of entanglement that can be 
defined on stabilizer states.  They found entanglement for small $n$, but 
for blocks above a certain size, they found nothing until they reached
the global entanglement scale.   Let us call the largest sub-global value of 
$n$ for which they found entanglement, $n_{{\rm{T}}}.$

By contrast, if we perform one iteration of a renormalization scheme such 
as that described above, we will obtain another toric code on a smaller, 
rescaled lattice.  This will also have short-range entanglement, by 
the result in \cite{ToricEnt}, and the iteration can be repeated until the 
logical qubit is decoded.  Thus {\emph{in this sense}} these codes can be 
said to have entanglement on all length scales.  It would seem to be 
essential to define this entanglement in terms of a renormalizaton procedure. 

However, both of these measures make sense and the number $n_{{\rm{T}}}$ 
does capture an important feature of this code.  Therefore, let us define the 
{\bf{cardinality}} of an entanglement type to be the number of parties it 
involves, and the {\bf{characteristic cardinality}} to be the largest 
cardinality present in whatever system we are currently looking at; in this 
case $n_{{\rm{T}}}.$
Let us incorporate the rescaling symmetry of the renormalization by defining 
the {\bf{scaling cardinality}} of entanglement of these codes to be 
$n_{{\rm{T}}},$ the characteristic cardinality of the local density matrix at 
each iteration.  These definitions may be extended to other cases in the 
obvious way. 

\bigskip

Note that while toric codes are usually defined on surfaces with non-trivial 
topology, there is a method for constructing them in planes while keeping 
the stabilizer generators local \cite{toricboundary,planarcodes}.  
Alternatively, one can define the code on a plane and have it encode no 
qubits.  These solutions would still have global entanglement, even though  
they don't encode any qubits.  They would correspond to globally pure systems 
with non-degenerate ground states.  It is an open question whether the 
definition of the quantum $\varepsilon$-memory support given in 
subsection VI.3 can be adapted to this case.  However, I suspect that 
the definition of such an empty ``toric'' code on the plane in terms of 
entirely local interactions would require a truly infinite plane with no 
boundaries, because these codes are defined topologically.  
Truly infinite planes do not occur in nature and so I suspect we would only 
ever see the ``unwrapped'' planar toric codes, as these can be defined for 
finite sized systems with boundaries in a sensible, local way.  The problem 
of defining the quantum $\varepsilon$-memory support for empty toric codes 
would therefore be moot. 

\bigskip

\noindent
{\bf{Solutions to addition RNGs}}

The original Kadanoff construction is often modified so that after a block 
has been mapped to one or more spins a {\emph{single}} new spin is added to 
the block from the rest of the lattice; the renormalization procedure is 
then repeated with what is essentially the original block instead of using 
a concatenated block structure.  This seems to have been done to try and 
keep the problem computationally feasible.  This seems to have become an 
almost universal practice, and indeed White's DMRG is a case in point. 

As hinted at above, this modification of the original Kadanoff construction 
may not be benign; there are classes of codes for which addition 
renormalization schemes may not capture the scaling behaviour of the 
entanglement.   

However, there are families of codes for which such an addition 
renormalization would be natural.  These are the quantum convolutional codes 
\cite{convcode1,convgood,convdesc}.  For an excellent introduction to these 
codes, see \cite{convrev}.  This features a modified version of the $5$-qubit 
code, adapted for use on a chain of qubits, such as a spin chain.   Note also 
that we could use the convolutional $5$-qubit code to make the stripes in the 
alternative ``brick'' tiling of the square lattice mentioned earlier, and 
shown in figure~\ref{brick}.  

Convolutional codes were developed to protect transmitted data (such as 
a radio broadcast) where the qubits pass through the device in a stream, and 
once the've gone, they are no longer accessible to processing.  It is also 
possible to use DFS-spaces in this situation, as shown in \cite{cryptoDFS}. 




\end{document}